\documentclass[a4paper,12pt]{article}
\usepackage[T1]{fontenc}
\usepackage[utf8]{inputenc}
\usepackage[english]{babel}
\usepackage{amsfonts}
\usepackage{enumitem}
\usepackage{amsmath}
\usepackage{amssymb}
\usepackage{graphicx}
\graphicspath{{immagini/}}
\usepackage{wrapfig}
\usepackage{tabularx}
\usepackage{booktabs}
\usepackage{float}
\usepackage[hidelinks]{hyperref}
\usepackage[italian]{varioref}
\usepackage{geometry}
\usepackage{textcomp}
\usepackage{multicol}
\usepackage{subfigure}
\usepackage{appendix}
\usepackage{listings}
\usepackage{url}
\usepackage{amsthm}
\usepackage{mathrsfs}

\usepackage[version=4]{mhchem}
\usepackage[dvipsnames,table]{xcolor}
\usepackage{verbatim}
\usepackage{fancyhdr}
\usepackage[square,numbers]{natbib}
\usepackage{csquotes}
\usepackage{parskip} 
\usepackage{bm}
\usepackage{empheq}
\usepackage{siunitx}
\usepackage{soul}
\usepackage{ulem}
\usepackage[version=4]{mhchem}
\DeclareMathAlphabet\mathbfcal{OMS}{cmsy}{b}{n}

\usepackage[T1]{fontenc}
\usepackage[utf8]{inputenc}
\usepackage[english]{babel}
\usepackage{amsfonts}
\usepackage{amsmath}
\usepackage{amssymb}
\usepackage{graphicx}
\graphicspath{{immagini/}}
\usepackage{wrapfig}
\usepackage{tabularx}
\usepackage{booktabs}
\usepackage{float}
\usepackage[hidelinks]{hyperref}
\usepackage[italian]{varioref}
\usepackage{geometry}
\usepackage{textcomp}
\usepackage{multicol}
\usepackage{subfigure}
\usepackage{appendix}
\usepackage{listings}
\usepackage{url}
\usepackage{amsthm}
\usepackage{mathrsfs}
\usepackage[version=4]{mhchem}
\usepackage[dvipsnames,table]{xcolor}
\usepackage{verbatim}
\usepackage{fancyhdr}
\usepackage[square,numbers]{natbib}
\usepackage{csquotes}
\usepackage{parskip} 
\usepackage{bm}
\usepackage{empheq}
\usepackage{siunitx}
\usepackage{soul}
\usepackage{ulem}
\usepackage[version=4]{mhchem}
\usepackage{titlesec}
\usepackage{afterpage}

\fancypagestyle{plain}{%
  \fancyhf{}%
  \fancyfoot{}%
}

\title{Tesi}
\author{Matilde Serena Illy}

\usepackage[T1]{fontenc}
\usepackage[utf8]{inputenc}
\usepackage{titlesec}

\titleformat{\section}[display]
  {\centering\normalfont\Large\bfseries}
  {\textit{Chapter \thesection}}
  {0pt}  {\titlerule[0.8pt]\vspace{1ex}\Huge}
  [{\vspace{1ex}\titlerule[0.8pt]\vspace{15mm}}]
  
\begin{document}

\titleformat{\subsection}{\Large\bfseries}{\thesubsection}{0.5em}{}
\titleformat{\subsubsection}{\large\bfseries}{\thesubsubsection}{0.5em}{}
\numberwithin{equation}{section}
\numberwithin{figure}{section}
\numberwithin{table}{section}

\begin{titlepage}
\begin{figure}[!htb]
    \centering
    \includegraphics[keepaspectratio=true,scale=1]{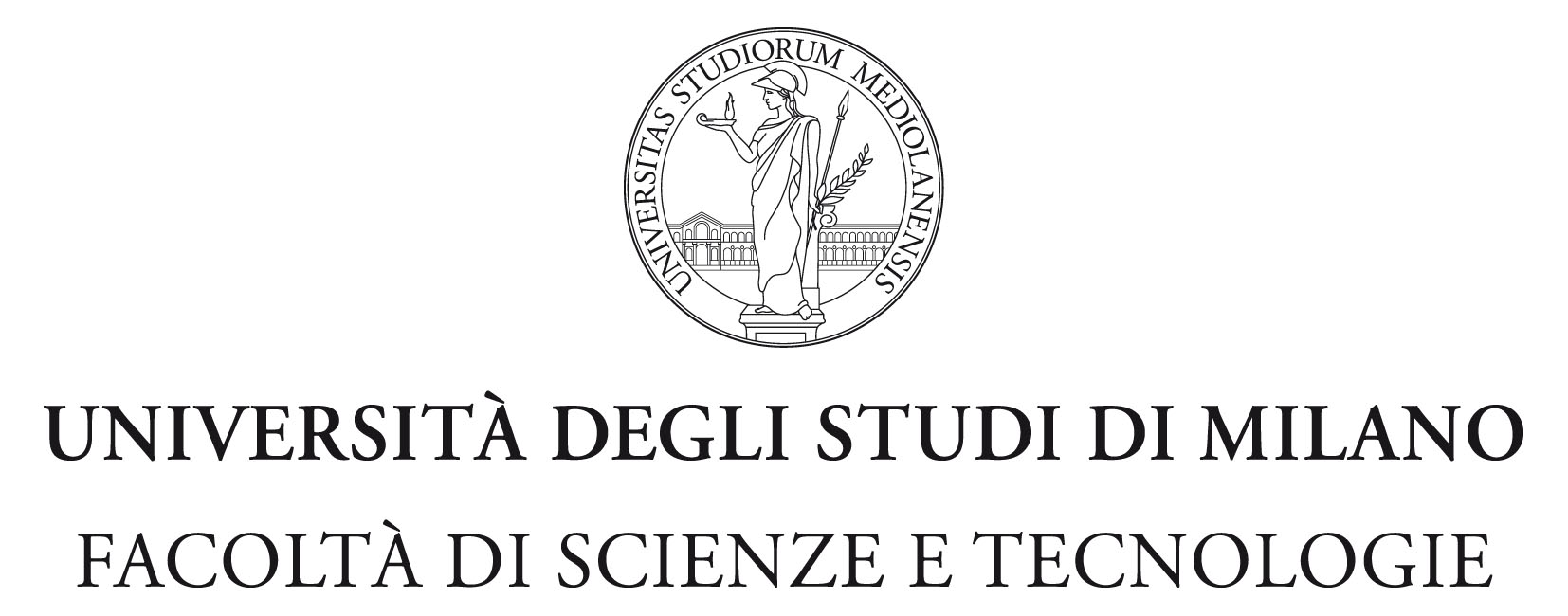}
\end{figure}

\begin{center}
    \LARGE{Corso di Laurea Triennale in Fisica}
\end{center}

\vspace{25mm}
\begin{center}
    {\LARGE{ \bf{Accelerated Reissner–Nordström \vspace{5mm} black hole in a swirling, magnetic universe}}}

\end{center}
\vspace{27mm}
\hfill
\normalsize
\begin{minipage}[t]{0.47\textwidth}
	{\Large{ \bf{Relatore:}}{\normalsize\vspace{3mm}
	\\\ \Large{Prof. Silke Klemm}}\vspace{5mm}
    \\\Large{ \bf{Relatore Esterno:}}{\normalsize\vspace{3mm}
	\\\ \Large{Dott. Marco Astorino}}}
\end{minipage}
\hfill
\begin{minipage}[t]{0.47\textwidth}\raggedleft\vspace{3cm}
	{\begin{flushleft}\Large{ \bf{Tesi di Laurea di:}}{\normalsize\vspace{2mm}
	\\ \Large{Matilde Serena Illy \vspace{2mm}\\Matr. 967268}} 
    \end{flushleft}}
\end{minipage}

\vspace{40mm}
\hrulefill
\\\centering{\large{ANNO ACCADEMICO 2022/2023}}

\end{titlepage}
\clearpage
\null
\thispagestyle{empty}
\clearpage


\pagenumbering{Roman}

\vspace*{\fill}
I declare that this thesis, partial duty for obtaining the bachelor degree in Physics at Università degli Studi Milano, is my own original work and, to the best of my knowledge and belief, it does not breach copyright or other intellectual property rights of a third party. \\

\begin{center}
    $\copyright$ Università degli Studi Milano.
\end{center}

\newpage

\tableofcontents
\begingroup
\let\clearpage\relax
\endgroup
\newpage
\cleardoublepage
\pagenumbering{arabic}

\addcontentsline{toc}{section}{Introduction}
\section*{Introduction}
In 1915, Albert Einstein published the theory of General Relativity: for the first time, gravity was no longer seen as a force but rather as a geometric manifestation of four-dimensional spacetime, mathematically modelled as a differentiable manifold with a lorent-\hspace{0.5cm}zian (signature $(-,+,+,+)$) metric tensor $g_{\mu\nu}$. Spacetime itself is described as curved or deformed by the matter and energy it contains: as John Wheeler famously said, ``Spacetime tells matter how to move, matter tells spacetime how to curve''. The mathematical formulation that embodies these concepts is known as Einstein's equations of General Relativity:

\hspace{6cm} $G_{\mu\nu}+\Lambda g_{\mu\nu}=8\pi T_{\mu\nu}$

where $G_{\mu\nu}$ is Einstein's tensor and $T_{\mu\nu}$ is the stress-energy tensor, which describes the distribution of matter and energy within the spacetime. $\Lambda$ is the cosmological constant, which will henceforth not be considered for reasons we will explain later on.

The Einstein equations are, in fact, 10 non-linear, second order partial differential equations for $g_{\mu\nu}$ in 4 variables. No general resolution methods are known, so in order to overcome this complexity and find solutions of physical interest, it can be assumed that a spacetime possesses particular symmetries. This concept made it possible to find, for instance, the Schwarzschild solution (1916), which describes an object that even Einstein did not believe could exist in nature: a (static, spherically symmetric) black hole.

Later on in the 20th century, many techniques were developed in order to generate solutions for these peculiar equations. In particular, we will focus on the Ernst generating technique (\cite{Ernst1}, \cite{Ernst2}): considering stationary and axially symmetric spacetimes in vacuum/electrovacuum conditions, it is possible to reformulate Einstein's equations in terms of two complex potentials, the ``Ernst potentials'', and thereby obtain a much simpler coupled system of two complex partial differential equations (PDEs) known as the Ernst equations. Ernst demonstrated this while considering the cosmological constant $\Lambda=0$ and later efforts to include it ultimately failed (\cite{cosmcost}). The most general form of a stationary and axisymmetric spacetime in electrovacuum conditions, known as the Lewis-Weyl-Papapetrou (LWP) metric, will be introduced and used in this thesis.

In a Lagrangian formulation, the Ernst equations can be found from an effective action. An analysis of the symmetries of the latter yields five different transformations that leave the equations formally unchanged. Therefore, by construction, by applying said transformations to a solution of the Ernst equations, a so-called ``seed'' metric, one obtains another solution. Three out of five are gauge symmetries, which can be reabsorbed by a coordinate transformation. Two of these gauge symmetries yield a physically equivalent spacetime but the remaining one is able to map a magnetic field into an electric field and viceversa without changing the overall spacetime’s structure. Instead, the so-called Ehlers and Harrrison transformations are not gauge symmetries and map a “seed” metric into a physically inequivalent spacetime.

It is then clear that we can simply apply these to a known solution in order to find a completely new one, without the need to solve a single PDE.

In this thesis, the composition of the Harrison and Ehlers transformations is applied to the Reissner-Nordström seed spacetime and the resulting metric and its physical and geometrical properties are analysed. Then, taking the accelerated Reissner-Nordström spacetime (charged C-metric) as seed, the same process is followed.

In the first chapter, stationary and axially symmetric metrics are introduced: these are the types of solutions that concern the Ernst generating technique. The Einstein-Maxwell equations are then rewritten in a simpler form, first given by Ernst in papers \cite{Ernst1} and \cite{Ernst2}. The second chapter aims to introduce the symmetries of the effective action, including the Ehlers and Harrison transformations. In the third chapter, the well-known Reissner-Nordström solution is presented and cast into LWP form so that chapter four can focus on the application of the Harrison-Ehlers hybrid transformation. The resulting solution, a Reissner-Nordström black hole embedded in a rotating, magnetic universe, is then obtained and its properties are analysed. Firstly, the focus is on ergoregions, then Dirac strings and conical singularities are explored. These two defects are found to affect the spacetime, therefore we look for a way to eliminate them. A new seed solution is introduced and cast into LWP form in chapter five: the accelerated Reissner-Nordström black hole. This is done because accelerated spacetimes (C-metrics) always present conical singularities, so the merging of two contributions to the nodal defects in the final spacetime may be a way to regularise it. Once the Harrison-Ehlers transformation is applied and the resulting metric is found, its physical properties and geometrical defects can finally be analysed. It is shown that the conical singularity can indeed be eliminated but, due to Dirac strings, the spacetime can never be fully regular at low energies.

Throughout this work I will use natural units $G=c=1$, the signature of the metric will be $(-,+,+,+)$ and the coordinate tetrad will have the following order: $(t,r,\phi,\theta)$.
\clearpage

\section[The Ernst solution generating technique]{\Large{The Ernst solution generating technique}}
In 1968, Frederick J. Ernst demonstrated that any axially symmetric and stationary solution of the Einstein-Maxwell equations can be described by two complex equations: the Ernst equations. In this chapter we will summarize the Ernst method and introduce the most general stationary and axisymmetric spacetime, the Lewis-Weyl-Papapetrou metric.

\subsection{The LWP metric}
The kind of spacetime that is of interest to us is axially symmetric and stationary: we are not limited to static, spherically symmetric objects like spheres but can explore more interesting shapes that rotate with a constant angular velocity. Let us define the properties of a spacetime that we can describe using Ernst's method.

If a spacetime admits a timelike Killing vector $\xi^\mu$ whose orbits are complete, it is said to be stationary. If instead it admits a spacelike Killing vector $\psi^\mu$ whose orbits are closed, then it is an axially symmetric spacetime.

A stationary and axisymmetric spacetime holds both these symmetries with the addition of commutativity between the two Killing vectors: [$\xi$,$\psi$]=0. This enables us to choose coordinates such as ($x^0$=t,$x^1$=$\phi$,$x^2$,$x^3$), in order that $\xi^\mu$=$\partial_t$ and $\psi^\mu$=$\partial_\phi$ are coordinate vector fields, which means that the metric components will be independent of t and $\phi$.

Our set of 10 equations is no longer in 4 coordinates, but 2. To further simplify metric components we may use the following theorem, whose proof can be found in \cite{Wald}.

\vspace{0.2cm}

\textbf{Theorem}
   
\textit{ Let $\xi^\mu$ and $\psi^\mu$ be two commuting Killing vector fields such that:}
\begin{enumerate}
    \item \textit{$\xi_{[\mu}$$\psi_\nu$$\nabla_\sigma$$\xi_{\lambda]}$ and $\xi_{[\mu}$$\psi_\nu$$\nabla_\sigma$$\psi_{\lambda]}$ each vanishes at, at least, one point of the spacetime.
    \item $\xi^\mu$$R_\mu^{[b}$$\xi^\sigma$$\psi^{\lambda]}$=$\psi^\mu$$R_\mu^{[b}$$\xi^\sigma$$\psi^{\lambda]}$=0.}
\end{enumerate}    
\textit{Then the 2-planes orthogonal to $\xi^\mu$ and $\psi^\mu$ are integrable.}

\vspace{0.2cm}

Condition \textit{1.} defines circular spacetimes and \textit{2.} is satisfied, trivially, when $R_{\mu\nu}$=0 everywhere, but also if $T_{\mu\nu}$ is the stress-energy tensor of a stationary and axisymmetric electromagnetic field, which we assume to be true.

This theorem has two main consequences: firstly, if the coordinate vectors of $t$ and $\phi$ generate a tangent space, then $x^2$ and $x^3$ will span the tangent space that is tangent to this one. This means that the mixed terms of the line elements are null, reducing the metric to six unknown functions rather than the ten we started with. At this point, if we rename certain line elements, the metric looks like this:

\begin{equation*} g_{\mu\nu} = \begin{pmatrix}
-V & W & 0 & 0 \\
W & X & 0 & 0 \\
0 & 0 & g_{22} & g_{23} \\
0 & 0 & g_{23} & g_{33}
\end{pmatrix}
\end{equation*}

where -V=-$\xi^\mu$$\xi_\mu$, W=$\xi^\mu$$\psi_\mu$ and X=$\psi^\mu$$\psi_\mu$.

Let us now choose the two remaining coordinates $x^2$ and $x^3$: assuming that $\nabla_\mu$$\rho$$\neq0$, we define $x^2$ as $\rho$ such that

\begin{equation} \label{eq:1.1} \rho^2=VX+W \end{equation}

which is minus the determinant of the $t-\phi$ component of the metric. In order to eliminate the $g_{23}$ component, we define $x^3$ as $z$ so that $\nabla_\mu \rho \perp \nabla_\mu$z, by setting z constant along $\nabla^\mu\rho$'s integral curves. Using these coordinates and defining $\omega=W/V$, we obtain:

\begin{equation*}
   ds^2=-V(dt-\omega d\phi)^2+\cfrac{1}{V}\hspace{0.1cm}\rho^2d\phi^2+\Omega^2(d\rho^2+\Lambda dz^2) 
\end{equation*}

which is the general form of a stationary and axisymmetric spacetime satisfying the theorem we wish to use.

We can further improve the metric if we consider a vacuum spacetime and set $R_{\mu\nu}=0$, which yields $D^\mu D_\mu \rho =0$\footnote{$D$ is the covariant derivative on the two-dimensional surface generated by $\rho$ and z.}. This implies that $\Lambda$ is a function on z only and, given the freedom we are granted in the definition of z, we can choose $\Lambda=1$ without loss of generality.

Finally, setting $V=f$ and $\gamma=\frac{1}{2} \ln(V\Omega^2)$, we obtain the Lewis-Weyl-Papapetrou (LWP) metric:

\begin{equation} \label{eq:1.2}
ds^2=-f(dt-\omega d\phi)^2+\cfrac{1}{f}\hspace{0.1cm}(e^{2\gamma}(d\rho^2+dz^2)+\rho^2d\phi^2)
\end{equation}

This is the most general metric in the vacuum and in the electrovacuum case (because $D^\mu D_\mu \rho =0$ still holds).

Aside from the fact that this is an exceptionally simple metric, we notice that all functions depend at most from both $\rho$ and z, whose differentials appear as they would in flat spacetime. For this reason, we can now use flat differential operators.

\subsection{The Ernst method}
We will now analyse Einstein's theory of General Relativity coupled with Maxwell's Electromagnetism. We can obtain the field equations by varying the following action, using the principle of least action:

\begin{equation} \label{eq:2.1} S(g_{\mu\nu},A_\mu)=\frac{1}{16\pi}\int (R-F^{\mu\nu}F_{\mu\nu})\sqrt{-g}\hspace{0.1cm} d^4x \end{equation}

where $F_{\mu\nu}$=$\partial_\mu A_\nu-\partial_\nu A_\mu$ is the Faraday tensor and $A=A_t(\rho,z)dt+A_\phi(\rho,z)d\phi$ \footnote{Due to the symmetries involved, $A$'s only non zero components are the ones corresponding to the ``Killing coordinates'' and they do not depend on said coordinates but only on $\rho$ and $z$.} is the most general one-form potential satisfying our symmetry conditions.

The Einstein equations for the metric may be written as follows:

\begin{equation*}
    G_{\mu\nu}=8\pi T_{\mu\nu}
\end{equation*}

An equivalent but more explicit expression is

\begin{equation} \label{eq:2.2} R_{\mu\nu}-\frac{1}{2}g_{\mu\nu} R=2\left(F_{\mu\alpha}F_\nu^\alpha-\frac{1}{4}g_{\mu\nu}F_{\alpha\beta}F^{\alpha\beta} \right) \end{equation}

where the stress-energy tensor $T_{\mu\nu}$ is expressed in terms of $F_{\mu\nu}$.

The Maxwell equations for $A_\mu$ may be written in a covariant form using the Faraday tensor:

\begin{equation} \label{eq:2.3} \nabla^\mu F_{\mu\nu}=\partial^\mu \sqrt{-g}F_{\mu\nu}=0 \end{equation}

Solving these equations gives us the functions present in the LWP metric (\ref{eq:1.2}) and the potential $A$, but these can also be obtained using the much less complicated Ernst's technique we shall now present.

From now on, the differential operators $\nabla$ and $\nabla^2$ will simply be the flat gradient and laplacian in cylindrical Weyl coordinates $(\rho,z,\phi)$.

In papers \cite{Ernst1} and \cite{Ernst2}, considering the Einstein-Maxwell field equations ((\ref{eq:2.2}), (\ref{eq:2.3})) in axially symmetric and stationary spacetimes, Ernst began the demonstration for his equations with the following Lagrangian density

\begin{equation} \begin{split} \label{eq:2.5}
L = & -\frac{1}{2}\rho f^{-2}\nabla f\cdot\nabla f+\frac{1}{2}\rho^{-1} f^{2}\nabla\omega\cdot\nabla\omega+2\rho f^{-1}A_t\nabla A_t\cdot\nabla A_t\ +\\ &-2\rho^{-1}f(\nabla A_\phi-\omega\nabla A_t)\cdot(\nabla A_\phi-\omega\nabla A_t)
\end{split} \end{equation}

which can be used to derive the field equations when considering a metric in the electric LWP form (\ref{eq:1.2}).

Through some manipulation of the Euler-Lagrange equations deriving from (\ref{eq:2.5}), we obtain two sets of equations:

\begin{equation} \label{eq:2.6} \nabla\cdot(\rho^{-2}f(\nabla A_\phi-\omega\nabla A_t))=0 \end{equation}

\begin{equation} \label{eq:2.7} \nabla\cdot(f^{-1}\nabla A_t+\rho^{-2}f\omega(\nabla A_\phi-\omega\nabla A_t))=0 \end{equation}

The above relations regard the electromagnetic potential, while

\begin{equation} \label{eq:2.8} \nabla\cdot(\rho^{-2}f^2\nabla\omega-4\rho^{-2}f A_t(\nabla A_\phi-\omega\nabla A_t))=0 \end{equation}

\begin{equation} \begin{split} \label{eq:2.9} f\nabla^2 f= & \nabla f\cdot\nabla f-\rho^{-2}f^4\nabla\omega\cdot\nabla\omega+2f\nabla A_t\cdot\nabla A_t+\\&+2\rho^{-2}f^3(\nabla A_\phi-\omega\nabla A_t)\cdot(\nabla A_\phi-\omega\nabla A_t) \end{split} \end{equation}

regard the gravitational potential.

What Ernst noticed is that, by defining a set of scalar functions, these equations can be simplified. In fact, two of them can be seen as integrability conditions for the existence of two new potentials: (\ref{eq:2.6}) helps us define $\tilde{A_\phi}$, while (\ref{eq:2.8}) helps us define $h$.

The divergence of the cross product of the azimuthal unit vector $\hat{\phi}$ and a function independent of $\phi$ (let us call it $\tilde{A_\phi}$) is zero. Therefore, we can define a magnetic scalar potential $\tilde{A_\phi}$, often referred to as ``twisted magnetic potential'', as follows:

\begin{equation} \label{eq:2.10} \hat{\phi} \times \nabla\tilde{A_\phi} = \rho^{-1}f(\nabla A_\phi-\omega\nabla A_t) \end{equation}

It follows that

\begin{equation*}
\rho^{-1}\hspace{0.1cm}\hat{\phi} \times \nabla A_\phi = -[f^{-1}\nabla \tilde{A_\phi}-\rho^{-1}\omega\hat{\phi}\nabla A_t] \end{equation*}

and therefore that

\begin{equation} \label{eq:2.11}
 \nabla[f^{-1}\nabla\tilde{A_\phi}-\rho^{-1}\omega\hspace{0.1cm}\hat{\phi}\times\nabla A_t]=0
 \end{equation}

On the other hand, (\ref{eq:2.7}) assumes the form

\begin{equation*}
 \nabla[f^{-1}\nabla A_t+\rho^{-1}\omega\hspace{0.1cm}\hat{\phi}\times\nabla\tilde{A_\phi}]=0
 \end{equation*}

Comparing the latter with (\ref{eq:2.11}), it appears useful to define the complex electromagnetic potential $\boldsymbol{\Phi}$: 

\begin{equation} \label{eq:2.12} \boldsymbol{\Phi}=A_t+i \tilde{A_\phi} \end{equation}

The new potential $\boldsymbol{\Phi}$ satisfies

\begin{equation} \label{eq:10}
 \nabla[f^{-1}\nabla\boldsymbol{\Phi}-i\rho^{-1}\omega\hspace{0.1cm}\hat{\phi}\times\nabla\boldsymbol{\Phi}]=0
 \end{equation}

which allows us to rewrite (\ref{eq:2.8}) in the following way:

\begin{equation*}
\nabla[\rho^{-2}f^2\nabla\omega-2\rho^{-1}\omega\hspace{0.1cm}\hat{\phi}\times Im(\boldsymbol{\Phi}^*\nabla\boldsymbol{\Phi})]=0
 \end{equation*}

This can be seen as an integrability condition for the existence of a new potential $h$, such that:

\begin{equation} \label{eq:2.13} \hat{\phi} \times \nabla h = \rho^{-1}f^2\nabla\omega-2\hspace{0.1cm}\hat{\phi}\times Im(\boldsymbol{\Phi}^*\nabla\boldsymbol{\Phi}) \end{equation}

It follows that

\begin{equation*}
\rho^{-1}\hspace{0.1cm}\hat{\phi} \times \nabla\omega = -f^{-2}[\nabla h+2Im(\boldsymbol{\Phi}^*\nabla\boldsymbol{\Phi})]
\end{equation*}

and hence that

\begin{equation} \label{eq:12}
 \nabla[f^{-2}[\nabla h+2Im(\boldsymbol{\Phi}^*\nabla\boldsymbol{\Phi})]]=0
 \end{equation}

Conversely, (\ref{eq:2.9}) can be rewritten as follows:

\begin{equation} \label{eq:13}
f\nabla^2 f=\nabla f\nabla f-[\nabla h+2Im(\boldsymbol{\Phi}^*\nabla\boldsymbol{\Phi})]\cdot[\nabla h+2Im(\boldsymbol{\Phi}^*\nabla\boldsymbol{\Phi})]+2f\nabla\boldsymbol{\Phi}\nabla\boldsymbol{\Phi}^*
\end{equation}

Comparing the two latter equations, we see that it is favourable to introduce the gravitational potential $\mathbfcal{E}$:

\begin{equation} \label{eq:2.14} \mathbfcal{E}=f-|\boldsymbol{\Phi}|^2+ih \end{equation}

Now, by combining (\ref{eq:12}) with (\ref{eq:13}) we obtain (\ref{eq:2.16}), while (\ref{eq:10}) yields (\ref{eq:2.17}). At last, the Ernst equations:

\begin{equation} \label{eq:2.16}
(Re(\mathbfcal{E})+|\boldsymbol{\Phi}|^2)\nabla^2\mathbfcal{E}=(\nabla\mathbfcal{E}+2\boldsymbol{\Phi}^*\nabla\boldsymbol{\Phi})\nabla\mathbfcal{E}
\end{equation}

\begin{equation} \label{eq:2.17}
(Re(\mathbfcal{E})+|\boldsymbol{\Phi}|^2)\nabla^2\boldsymbol{\Phi}=(\nabla\mathbfcal{E}+2\boldsymbol{\Phi}^*\nabla\boldsymbol{\Phi})\nabla\boldsymbol{\Phi}
\end{equation}

This coupled system of complex vectorial differential equations effectively substitutes the Einstein-Maxwell field equations that derive from action (\ref{eq:2.1}) when dealing with an axisymmetric stationary spacetime. They can therefore be used to retrieve the $f$ and $\omega$ functions of the LWP metric. There is, however, no mention of $\gamma$, which is present in (\ref{eq:2.1}) and can be obtained from the corresponding field equations. The first order differential equations, uncoupled with the above system, responsible for $\gamma$'s behaviour are:

\begin{equation} \begin{split} \label{eq:2.18}
    \partial_\rho\gamma(\rho,z)=&\frac{\rho}{4(Re(\mathbfcal{E})+|\boldsymbol{\Phi}|^2)^2}[(\partial_\rho\mathbfcal{E}+2\boldsymbol{\Phi}^*\partial_\rho\boldsymbol{\Phi})(\partial_\rho\mathbfcal{E}^*+2\boldsymbol{\Phi}\partial_\rho\boldsymbol{\Phi}^*)-(\partial_z\mathbfcal{E}+2\boldsymbol{\Phi}^*\partial_z\boldsymbol{\Phi})\cdot\\&\cdot(\partial_z\mathbfcal{E}^*+2\boldsymbol{\Phi}\partial_z\boldsymbol{\Phi}^*)]
    -\frac{\rho}{Re(\mathbfcal{E})+|\boldsymbol{\Phi}|^2}(\partial_\rho\boldsymbol{\Phi}\partial_\rho\boldsymbol{\Phi}^*-\partial_z\boldsymbol{\Phi}\partial_z\boldsymbol{\Phi}^*)
\end{split} \end{equation}

\begin{equation} \begin{split} \label{eq:2.19}
\partial_z\gamma(\rho,z)=&\frac{\rho}{4(Re(\mathbfcal{E})+|\boldsymbol{\Phi}|^2)^2}[(\partial_\rho\mathbfcal{E}+2\boldsymbol{\Phi}^*\partial_\rho\boldsymbol{\Phi})(\partial_z\mathbfcal{E}^*+2\boldsymbol{\Phi}\partial_z\boldsymbol{\Phi}^*)+(\partial_z\mathbfcal{E}+2\boldsymbol{\Phi}^*\partial_z\boldsymbol{\Phi})\cdot\\&\cdot(\partial_\rho\mathbfcal{E}^*+2\boldsymbol{\Phi}\partial_\rho\boldsymbol{\Phi}^*)]-\frac{\rho}{Re(\mathbfcal{E})+|\boldsymbol{\Phi}|^2}(\partial_\rho\boldsymbol{\Phi}\partial_z\boldsymbol{\Phi}^*+\partial_z\boldsymbol{\Phi}\partial_\rho\boldsymbol{\Phi}^*)
\end{split} \end{equation}

\clearpage

\section{The Symmetry Group}
In the nineteenth century, mathematician Sophus Lie introduced the analysis of continuous "Lie" transformation groups, which consist in transformations that leave an ordinary or partial differential equation unchanged. As a consequence, they map the solution set of a given system into itself, allowing us to find a new solution from an existing one. This method is exploited by the Ernst generating technique, which finds new solutions to the Einstein-Maxwell field equations starting from a "seed" solution, using the symmetries of the Ernst equations. If the equations of motion are left unchanged under a given transformation, we can then apply it to a known solution and find another one. For the purposes of this thesis, explaining the origin of these symmetries is not relevant but an extensive description can be found in \cite{Diffeq}.

\subsection{Symmetries of the Effective Action}
Let us consider the following effective action:

\begin{equation*}
S(\mathbfcal{E},\boldsymbol{\Phi})=\int \left[\frac{(\nabla\mathbfcal{E}+2\boldsymbol{\Phi}^*\nabla\boldsymbol{\Phi})(\nabla\mathbfcal{E}^*+2\boldsymbol{\Phi}\nabla\boldsymbol{\Phi}^*)}{(\mathbfcal{E}+\mathbfcal{E}^*+2|\boldsymbol{\Phi}|^2)^2}-2\frac{\nabla\boldsymbol{\Phi}\nabla\boldsymbol{\Phi}^*}{\mathbfcal{E}+\mathbfcal{E}^*+2|\boldsymbol{\Phi}|^2} \right] \rho d\rho dz
\end{equation*}

It is possible to demonstrate that $S(\mathbfcal{E},\boldsymbol{\Phi})$ and its equations of motions remain unchanged under the action of the following independent transformations\footnote{For instance, this is shown in \cite{Mart}.}:
    
\begin{enumerate} [label=\Roman*.]

\item $\mathbfcal{E}\longrightarrow\mathbfcal{E}'=\lambda\lambda^*\mathbfcal{E}$,\hspace{1.5cm} $\boldsymbol{\Phi}\longrightarrow\boldsymbol{\Phi}'=\lambda\boldsymbol{\Phi}$

\item
$\mathbfcal{E}\longrightarrow\mathbfcal{E}'=\mathbfcal{E}+ib$ ,\hspace{1.5cm} $\boldsymbol{\Phi}\longrightarrow\boldsymbol{\Phi}'=\boldsymbol{\Phi}$

\item $\mathbfcal{E}\longrightarrow\mathbfcal{E}'=\cfrac{\mathbfcal{E}}{1+ij\mathbfcal{E}}$ ,\hspace{1.5cm} $\boldsymbol{\Phi}\longrightarrow\boldsymbol{\Phi}'=\cfrac{\boldsymbol{\Phi}}{1+ij\mathbfcal{E}}$

\item
$\mathbfcal{E}\longrightarrow\mathbfcal{E}'=\mathbfcal{E}-2\beta^*\boldsymbol{\Phi}-\beta\beta^*$ ,\hspace{1.5cm} $\boldsymbol{\Phi}\longrightarrow\boldsymbol{\Phi}'=\boldsymbol{\Phi}+\beta$

\item 
$\mathbfcal{E}\longrightarrow\mathbfcal{E}'=\cfrac{\mathbfcal{E}}{1+2\alpha^*\boldsymbol{\Phi}-\alpha\alpha^*\mathbfcal{E}}$ ,\hspace{1.5cm} $\boldsymbol{\Phi}\longrightarrow\boldsymbol{\Phi}'=\cfrac{\alpha\mathbfcal{E}+\boldsymbol{\Phi}}{1+2\alpha^*\boldsymbol{\Phi}-\alpha\alpha^*\mathbfcal{E}}$

\end{enumerate}

with $\lambda,\alpha$ and $\beta$ $\in$ $\mathbb{C}$ and $b,j$ $\in$ $\mathbb{R}$. This is a group with 8 real parameters, namely $SU(1,2)$.

It is important to note that not all of these will map our seed metric into a physically inequivalent new solution, in fact transformations II and IV are gauge transformations, which means that they yield the same metric but in another coordinate system (hence a coordinate transformation can eliminate their effect). I is also a gauge transformation, however given a spacetime in electrovacuum conditions, it can map a magnetic field into an electric field and viceversa without changing the overall spacetime's structure.  Transformations III and V are not gauge symmetries and they are nontrivial: III, known as the Ehlers transformation, adds a NUT or gravomagnetic parameter $j$ (analogous to the magnetic charge in electromagnetism) to the metric; while V, known as the Harrison transformation, adds the parameter $\alpha$ which provides an electromagnetic field.

For example, applying the Ehlers transformation to the Schwarzschild metric written in magnetic LWP form (see Section 3.2) yields a Schwarzschild black hole immersed in a rotating universe, as described in \cite{Swirl}. Instead, applying the Ehlers transformation to the Schwarzschild metric in electric LWP form yields a Taub-NUT sort of metric, as shown in \cite{Mart}.

One last interesting remark: these transformations are continuous, but we can also have discrete ones. For instance, the inversion transformation can be obtained by $I\cdot III\cdot II$ (\cite{enhanced}), where $\cdot$ indicates function composition, and the result is the following:

\begin{equation*} \begin{cases}
    \mathbfcal{E}\longrightarrow\mathbfcal{E}'=\frac{1}{\mathbfcal{E}} \\
    \boldsymbol{\Phi}\longrightarrow\boldsymbol{\Phi}'=\frac{\boldsymbol{\Phi}}{\mathbfcal{E}}
\end{cases} \end{equation*}

\clearpage

\section{The Reissner-Nordström solution}
The Reissner-Nordström metric is used in General Relativity to describe the spacetime around a spherically symmetric, static black hole that features an electric charge. It was discovered independently by Reissner, Nordström, Weyl and Jeffery between 1916 and 1921. In this chapter we will present this and cast it in the LWP form, as it will serve as our seed metric.

\subsection{The solution}
The Reissner-Nordström metric can be written as follows in spherical coordinates ($t,r,\phi,\theta$):

\begin{equation} \label{eq:3.0}
    ds^2=-Q(r)dt^2+\frac{dr^2}{Q(r)}+r^2d\theta^2+r^2\sin^2\theta\hspace{0.1cm} d\phi^2
\end{equation}

with

\begin{equation*}
Q(r)=1-\cfrac{2m}{r}+\cfrac{e^2}{r^2}
\end{equation*}

$e$ and $m$ are respectively the electrical charge and the mass of the black hole. The parameters e and m have the constraint that $e<m$ because black holes with a greater charge than mass cannot exist in nature, otherwise the event horizon would not exist. The electromagnetic vector potential compatible with this metric is $A=(-\frac{e}{r}, 0, 0, 0)$.

We can see that this is a very simple metric, in fact we obtain the Schwarzschild solution in the limit $e\xrightarrow{}0$. It has been proven that this is the only spherically symmetric, asymptotically flat and static (due to Birkhoff's theorem) solution to the Einstein-Maxwell field equations.

We also recall that the LWP metric was in cylindrical Weyl coordinates ($t,\rho,z,\phi$) while this is expressed in spherical ones. We can then use the following transformation to change coordinates:

\begin{equation} \label{eq:3.1}
\begin{cases}
\rho(r,\theta)=r\sin\theta \sqrt{Q} \\
z(r,\theta)=(r-m)\cos\theta
\end{cases}
\end{equation}

while $t$ and $\phi$ remain unchanged. Let us note that, while $\rho$ can be found by simply comparing the $dt^2$ components of the LWP metric with those of the Reissner-Nordström, finding $z$ is somewhat more complicated: since we don't yet know $\gamma$'s expression, we cannot exploit the $dr$ and $d\theta$ components.

What we instead do is solve a differential equation for $z$, knowing that the Reissner-Nordström metric does not contain a mixed $drd\theta$ term. We start with the assumption that $z(r,\theta$) is a separable function, i.e. $z(r,\theta)=f(r)g(\theta)$. We can now write the squared differential $dz^2$ in terms of $f$, $g$ and their (total) derivatives: this expression will have a mixed $drd\theta$ term and the trick is to equal this to minus the mixed term found in $d\rho^2$ (which is easy to calculate since we just found $\rho(r,\theta)$). We are now left with a separable PDE, so let's bring all functions dependent on $r$ only on the left-hand side and the $\theta$-dependent functions on the right-hand side:

\[\frac{g(\theta)g'(\theta)}{\cos(\theta)\sin(\theta)}=-\frac{r}{f(r)f'(r)}\left(Q(r)+\frac{m}{r}-\frac{e^2}{r^2}\right)=C \]

This is equal to a constant C because the left-hand side is a function of $\theta$ only, while the right-hand side is a function of $r$ only. We now have two ODEs (ordinary differential equations), which we can easily solve through integration and an appropriate choice of integration constants, and finally find the expression for $z(r,\theta$).

The coordinate change found can be used to compute the gradient operator: we will not need its full expression because in the equations we will use, this operator is found on both sides, making common terms trivial. In addition, only the derivatives with respect to $r$ and $\theta$ are relevant, because the metric does not depend on its Killing coordinates. We will therefore use the following expression:

\begin{equation}
\nabla\propto\hat{e_r}\sqrt{\Delta(r)}\partial_r+\hat{e_\theta}\hspace{0.1cm}\partial_\theta
\end{equation}

with $\Delta(r)=r^2\cdot Q(r)$.

For simplicity, we report the $d\rho^2+dz^2$ component of the LWP metric in spherical coordinates:

\[ d\rho^2+dz^2=(r^2-2mr+e^2\cos^2\theta+m^2\sin^2\theta)\left(\frac{dr^2}{r^2-2mr+e^2}+d\theta^2 \right) \]

\subsection{Reissner-Nordström in magnetic LWP form}
Let us recall the LWP metric we introduced in the previous chapter:

\[ds^2=-f(dt-\omega(r,\theta) d\phi)^2+\frac{1}{f}(e^{2\gamma}(d\rho^2+dz^2)+\rho^2d\phi^2) \]

This is called "electric" LWP metric. If we apply a transformation called a double Wick rotation (an inversion of space and azimuth), namely

\begin{equation} \begin{cases} \label{eq:3.2}
t\xrightarrow{} i\psi \\
\phi\xrightarrow{} i\tau
\end{cases} 
\end{equation}

we obtain, from the electric LWP metric, the magnetic version of this metric:

\begin{equation} \label{eq:3.3}
    ds^2=-f(d\phi-\omega(r,\theta) dt)^2+\frac{1}{f}(\rho^2d\phi^2-e^{2\gamma}(d\rho^2+dz^2))
\end{equation}

where we implicitly applied the trivial conversion $\psi\xrightarrow{}\phi$, $\tau\xrightarrow{} t$ for simplicity.

This is the version of the LWP that we will now compare with the Reissner-Nordström metric (\ref{eq:3.0}).

As a first step, it is straightforward to notice that $\omega(r,\theta)=0$, since the Reissner-Nordström solution is non rotating (static). Comparing then the $d\phi^2$ and $d\rho^2+dz^2$ terms, we obtain:

\begin{equation} \begin{cases} \label{eq:3.4}
f(r,\theta)=-r^2\sin^2\theta \\
e^{2\gamma(r,\theta)}=\cfrac{r^4\sin^2\theta}{r^2-2mr+e^2\cos^2\theta+m^2\sin^2\theta}
\end{cases} 
\end{equation}

We are not done just yet: we still need to find the Ernst potentials $\mathbfcal{E}$ and $\boldsymbol{\Phi}$. In the magnetic LWP case, their expressions only slightly differ from (\ref{eq:2.11}), (\ref{eq:2.13}):

\begin{equation} \begin{cases} \label{eq:3.5}
\boldsymbol{\Phi}=A_\phi+i\tilde{A_t} \\
\mathbfcal{E}=f+ih-|\boldsymbol{\Phi}|^2
\end{cases} 
\end{equation}

In order to find $\boldsymbol{\Phi}$, we need the expressions for $A_\phi$ and $\tilde{A_t}$, so we can use the following differential equation\footnote{Throughout this thesis we will use the orthonormal basis ($\hat{e_r}, \hat{e_\phi}, \hat{e_\theta}$) for all vectorial operations.}:

\begin{equation} \label{eq:3.6}
\nabla \tilde{A_t}=\frac{f}{\rho}\hat{e_\phi}\times(\nabla A_t+\omega\nabla A_\phi)
\end{equation}

Let us rewrite this equation in its vectorial components for simplicity. We obtain:

\begin{equation*}
\sqrt{\Delta(r)}\hspace{0.1cm}\partial_r\tilde{A_t}=\frac{f}{\rho}\left(\partial_\theta A_t+\omega\hspace{0.1cm}\partial_\theta A_\phi\right)
\end{equation*}

along $\hat{e_r}$, and

\begin{equation*}
\partial_\theta\tilde{A_t}=-\frac{f}{\rho}\sqrt{\Delta(r)}\left(\partial_r A_t+\omega\hspace{0.1cm}\partial_r A_\phi \right)
\end{equation*}

along $\hat{e_\theta}$.

We know that $A_t=-\cfrac{e}{r}$ \hspace{0.05cm} and $A_\phi=0$, so we find $\tilde{A_t}=-e\cos\theta$. Therefore

\begin{equation} \label{eq:3.7}
\boldsymbol{\Phi}(\theta)=-ie\cos\theta
\end{equation}

In our search for $\mathbfcal{E}$, we need to find $h(r,\theta)$'s expression. We do so by using

\begin{equation} \label{eq:3.8}
\nabla h=\frac{f^2}{\rho}\hat{e_\phi}\times\nabla\omega-2Im(\boldsymbol{\Phi}^*\nabla\boldsymbol{\Phi})
\end{equation}

Written in vectorial components, this becomes:

\begin{equation*}
\sqrt{\Delta(r)}\hspace{0.1cm}\partial_r h=\frac{f^2}{\rho}\partial_\theta \omega -2\sqrt{\Delta(r)}(A_\phi \partial_r \tilde{A_t}+\tilde{A_t}\partial_r A_\phi) 
\end{equation*}

along $\hat{e_r}$, and

\begin{equation*}
\partial_\theta h=-\frac{f^2}{\rho}\sqrt{\Delta(r)}\hspace{0.1cm}\partial_r \omega -2(A_\phi \partial_\theta \tilde{A_t}+\tilde{A_t}\partial_\theta A_\phi) 
\end{equation*}

along $\hat{e_\theta}$.

We know that $\omega(r,\theta)=0$ and $\boldsymbol{\Phi}^*\nabla\boldsymbol{\Phi}\in\mathbb{R}$: from this we deduce that $h(r,\theta)$ is at most a constant, so we can reabsorb it through a trivial coordinate change. We can therefore set $h(r,\theta)=0$. We now have an expression for Ernst's gravitational potential:

\begin{equation} \label{eq:3.9}
\mathbfcal{E}(r,\theta)=-r^2\sin^2\theta-e^2\cos^2\theta
\end{equation}

We have all the information we need about our seed metric and can, at last, present the transformation we will apply to it.
\clearpage

\section{The new solution}
In this chapter we will apply a ``hybrid'' transformation to our seed metric: the Harrison-Ehlers transformation. This will add not one but two parameters to the seed metric, which has some interesting implications.

\subsection{Finding the solution}
The hybrid transformation we mentioned is the composition of the Ehlers and Harrison transformations, which we rewrite here for convenience. 

\begin{equation*}
\begin{cases}    
 \mathbfcal{E}\longrightarrow\mathbfcal{E}'=\cfrac{\mathbfcal{E}}{1+ij\mathbfcal{E}} \hspace{2.3cm} \boldsymbol{\Phi}\longrightarrow\boldsymbol{\Phi}'=\cfrac{\boldsymbol{\Phi}}{1+ij\mathbfcal{E}} \\
 \mathbfcal{E}\longrightarrow\mathbfcal{E}'=\cfrac{\mathbfcal{E}}{1+B\boldsymbol{\Phi}-\cfrac{B^2}{4}\mathbfcal{E}} \hspace{1cm} \boldsymbol{\Phi}\longrightarrow\boldsymbol{\Phi}'=\cfrac{\boldsymbol{\Phi}-\cfrac{B}{2}\mathbfcal{E}}{1+B\boldsymbol{\Phi}-\cfrac{B^2}{4}\mathbfcal{E}}
 
\end{cases}
\end{equation*}

where we chose $\alpha=-\frac{B}{2}$, $B$ $\in$ $\mathbb{R}$, for the Harrison.

It turns out that the commutator of these two transformations is zero (\cite{acctype1}), which means that (Harrison)$\cdot$(Ehlers)=(Ehlers)$\cdot$(Harrison), where $\cdot$ indicates function composition. Therefore, our transformation is the following:

\begin{equation} \begin{cases} \label{eq:4.1}
  \boldsymbol{\Phi}'=\cfrac{\boldsymbol{\Phi}-\cfrac{B}{2}\mathbfcal{E}}{1+ij\mathbfcal{E}+B\boldsymbol{\Phi}-\cfrac{B^2}{4}\mathbfcal{E}} \\
 \mathbfcal{E}'=\cfrac{\mathbfcal{E}}{1+ij\mathbfcal{E}+B\boldsymbol{\Phi}-\cfrac{B^2}{4}\mathbfcal{E}}
\end{cases} \end{equation}

If we now remember the definition of these two potentials (\ref{eq:3.5}) and our expressions for $\mathbfcal{E}$ and $\boldsymbol{\Phi}$ ((\ref{eq:3.7}), (\ref{eq:3.9})), we can easily find the following functions:

\begin{equation} \label{eq:4.2}
    A_\phi'(r,\theta)=\frac{1}{|\Lambda(r,\theta)|^2}\left(e^2B\cos^2\theta-\frac{B}{2}\mathbfcal{E}\left(1-\frac{B^2}{4}\mathbfcal{E} \right)-je\mathbfcal{E} \cos\theta \right)
\end{equation}

\begin{equation} \label{eq:4.3}
    \tilde{A_t}'(r,\theta)=\frac{1}{|\Lambda(r,\theta)|^2}\left(j\frac{B}{2}\mathbfcal{E}^2-e\cos\theta\left(1+\frac{B^2}{4}\mathbfcal{E}\right)\right)
\end{equation}

\begin{equation} \label{eq:4.4}
    h'(r,\theta)=\frac{Be\cos\theta\mathbfcal{E}-j\mathbfcal{E}^2}{|\Lambda(r,\theta)|^2}
\end{equation}

\begin{equation} \label{eq:4.5}
    f'(r,\theta)=\frac{-r^2\sin^2\theta}{|\Lambda(r,\theta)|^2}=\frac{f(r,\theta)}{|\Lambda(r,\theta)|^2}
\end{equation}

To simplify the expressions, we did not write $\mathbfcal{E}(r,\theta)$ explicitly and named (\ref{eq:4.1})'s denominator $\Lambda(r,\theta)$.

By substituting $\mathbfcal{E}'$ in equations (\ref{eq:2.18}), (\ref{eq:2.19}), it can be easily verified that $\gamma'(r,\theta)=\gamma(r,\theta)$. Therefore, the only functions left to find are $\omega'(r,\theta)$ and $A_t'(r,\theta)$: this is done by using (\ref{eq:3.6}) and (\ref{eq:3.8}) respectively. The resulting expressions are:

\begin{equation} \label{eq:4.6}
    A_t'(r,\theta)=S(r,\theta)-\omega(r,\theta)A_\phi(r,\theta)
\end{equation}

where the auxiliary function $S(r,\theta)$ is

\begin{equation*}
    S(r,\theta)=\frac{e}{4r}\left(4-3B^2(r^2+(e^2+r(r-2m)) \cos^2\theta) \right)
\end{equation*}

and $\omega'(r,\theta)$ is

\begin{equation} \label{eq:4.7}
    \omega'(r,\theta)=\frac{B^3er^2-4Be+\cos\theta(e^2+r(r-2m))(B^3e \cos\theta-8j)}{2r}
\end{equation}

Interestingly, $\omega'(r,\theta)$ depends both on $B$ and $j$, which means that both the Ehlers and the Harrison transformations individually contribute to the rotation.

Let us elaborate on this peculiarity: the Ehlers transformation embeds the seed in a swirling universe, i.e. a rotating background; naturally, this initiates a rotation of the spacetime. Furthermore, since we chose $\alpha=-\frac{B}{2}$ $\in$ $\mathbb{R}$ as the Harrison parameter, we are embedding an electrically charged spacetime (Reissner-Nordström) in a magnetically charged background. For this reason, an $\Bar{E}\times\Bar{B}$ term (where $\Bar{E}$ is the electric field and $\Bar{B}$ is the magnetic field) appears in the stress-energy tensor, thereby generating the twist potential $\tilde{A_t}$ and contributing to the rotation. This is a typical scenario when considering magnetised black holes: an intrinsic electric charge coupled with an external magnetic field (and viceversa) yields this situation, as Ernst himself explained in \cite{Ernst3}.

For completeness, let us specify that the Harrison transformation with parameter $\alpha$ $\in$ $\mathbb{R}$, $\mathbb{I}$, $\mathbb{C}$ embeds the seed respectively in a magnetic, electric and electromagnetic background.

Now, let us insert all of the primed functions in the LWP metric in spherical coordinates in order to find the new solution:

\begin{equation} \label{eq:4.8}
    ds^2=-\frac{f(r,\theta)}{|\Lambda(r,\theta)|^2}(d\phi-\omega'(r,\theta) dt)^2+|\Lambda(r,\theta)|^2\left(Q(r)dt^2-\frac{dr^2}{Q(r)}-r^2d\theta^2 \right)
\end{equation}

\subsection{Physical and geometrical properties}
As we can see, (\ref{eq:4.8}) is a four parameter metric: the mass $m$, the electric charge of the seed $e$, the magnetic field of the background $B$ and the angular velocity of the background $j$. By construction, it is a solution of the Maxwell-Einstein field equations, but this has also been verified on the Wolfram Mathematica software.

The metric represents a non-asymptotically flat, non-diagonal deformation of the Reissner-Nordström black hole, of which it retains the causal structure. Indeed, it may be viewed as a Reissner-Nordström black hole inserted in a ``swirling'', magnetic universe. Let us discuss a hypothesis for the origin of the background rotation and magnetic field.

In their paper \cite{Emparan}, Emparan and Gutperle demonstrated that the Melvin universe, i.e. our magnetic background, can be interpreted as two infinitely distant (along the symmetry axis) Reissner-Nordström black holes at infinity with opposite magnetic charges. The rotating background, as explained in \cite{Removal}, can instead be thought to be caused by as a couple of infinitely faraway counter-rotating sources (at infinity), such as two black holes with opposite NUT charges or two counter-rotating galaxies. The latter interpretation has not been demonstrated, however it seems to be a valid hypothesis given the many analogies between the Melvin and swirling universes. In particular, they can both be obtained through a double Wick rotation (\ref{eq:3.2}) of spacetimes with a flat base manifold, respectively Reissner-Nordström and Taub–NUT (their extremal versions, where the mass equals the electric or NUT charge, so that the inner and outer event horizons coincide), as demonstrated in \cite{Swirl}.

\begin{figure}[H]
    \centering
    \includegraphics[width=0.50 \textwidth]{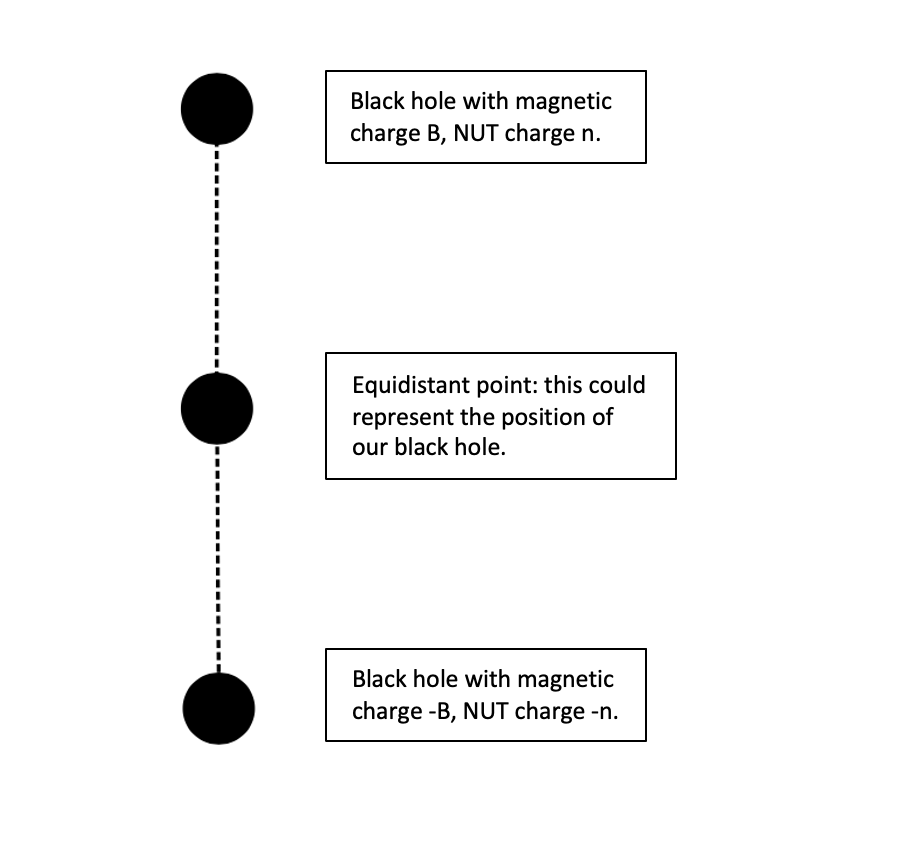}
    \caption{Schematic representation of the origin of our spacetime's background rotation and magnetic field. The two sources are infinitely distant, with axial symmetry, from the central black hole. }
    \label{fig:Disegno1}
\end{figure}

If we combine these interpretations, as sketched in Figure \ref{fig:Disegno1}, we have two black holes with opposite magnetic and NUT charges that lie infinitely far apart from each other and from the black hole we are describing. If we imagine the latter to be an equidistant point for the two black holes at infinity, then their presence can be thought to explain the gravitational whirlpool and magnetic field in the background.

Before moving on to other properties, it is useful to analyse the singularities that this spacetime presents. The Riemann tensor $R_{abcd}$ measures curvature but, because its components are coordinate dependent, it is hard to understand when it actually becomes infinite. However, it is possible to construct scalar quantities using $R_{abcd}$ and these are instead coordinate-independent, so their divergence is meaningful. The commonly used Kretschmann scalar $R^{abcd}R_{abcd}$, for instance, can be used to determine singularities of the curvature because it goes to infinity when it approaches one. For spacetime (\ref{eq:4.8}) and for a reasonable set of parameters ($m$, $e$, $j$, $B$), the Kretschmann scalar's diverges as $r\xrightarrow{}0$:

\begin{equation*}
    R^{abcd}R_{abcd}\approx \frac{1}{r^8}
\end{equation*}

meaning that $r=0$ is a singularity of curvature. Instead, the inner and outer event horizons $r=m\pm\sqrt{m^2-e^2}$ are coordinate singularities.

\subsubsection[Analysing the spacetime's frame-dragging]{\large{Analysing the spacetime's frame-dragging}}

This metric presents an angular velocity $\Omega(r,\theta)$ that does not remain constant along the z axis, as a matter of fact we have two different behaviours for $\theta=0$ and $\theta=\pi$:

\begin{equation*}
    \Omega|_{\theta=0}=\lim_{\theta\to 0}\left(-\frac{g_{t\phi}}{g_{\phi\phi}}\right)=\frac{-4Be-8j(e^2+r(-2m+r))+B^3e(e^2+2r(-m+r))}{2r}
\end{equation*}

\begin{equation*}
    \Omega|_{\theta=\pi}=\lim_{\theta\to \pi}\left(-\frac{g_{t\phi}}{g_{\phi\phi}}\right)=\frac{-4 B e + B^3 e r^2 + (B^3 e + 8 j) (e^2 + r (-2 m + r))}{2r}
\end{equation*}

Unlike what takes place in asymptotically flat spacetimes, the angular velocity $\Omega(r,\theta)$ is not constant on the symmetry axis. This feature is also present in the magnetised Reissner-Nordström (our solution but with the constraint $j=0$) and the magnetised Kerr solutions, as we can see in \cite{Gibb}.

The magnetised, swirling Reissner-Nordström solution represents the spacetime around a massive rotating object, which generates a rotational frame dragging known as the Lense-Thirring effect (this is explained in depth in \cite{Gravitation}). This effect causes the dragging of inertial frames and one of its consequences is that a distant observer will see that light travelling in the same direction as the object's rotation will move past it faster than light travelling in the opposite direction. In our case, the frame dragging is given by

\begin{equation} \label{eq:angv}
    \frac{d\phi}{dt}=-\frac{g_{t\phi}}{g_{\phi\phi}}=-\omega'(r,\theta)
\end{equation}

where $\omega'(r,\theta)$ is our metric function (\ref{eq:4.7}). 

If we consider the $r\xrightarrow{}\infty$ limit of our angular velocity (\ref{eq:angv}), we have:

\begin{equation*}
    \lim_{r\to \infty}\frac{d\phi}{dt}=\left(\frac{1}{2}B^3e(1+\cos^2\theta) -4j\cos\theta \right)r
\end{equation*}

Clearly, it grows unbounded in this limit thereby exceeding 1, i.e. the speed of light in our physical units. Does this imply the existence of superluminal observers in our spacetime and therefore the violation of the causality principle? To answer this, let's introduce CTCs.

Closed timelike curves are trajectories that return to their initial point in spacetime, violating the laws of Special Relativity and hence introducing paradoxes and the violation of the causality principle. Their existence is famously allowed in the Gödel spacetime (1949), a fact that deeply worried Einstein.

If CTCs do occur in our spacetime, then the answer to our previous question might be yes. To check this, we consider curves in which $t$, $r$ and $\theta$ remain constant within (\ref{eq:4.8}):

\begin{equation*}
    ds^2|_{t,r,\theta}=-\frac{f(r,\theta)}{|\Lambda|^2}d\phi^2=\frac{r^2\sin^2\theta}{|\Lambda|^2}d\phi^2
\end{equation*}

This expression is always positive, which means that the curves $ds^2|_{t,r,\theta}$ are always spacelike. Therefore, the occurrence of CTCs in our spacetime (\ref{eq:4.8}) can be excluded. Then, the fact that our angular velocity exceeds 1 for $r\xrightarrow{}\infty$ and the consequent violation of causality can simply be explained by a ``bad'' choice of coordinates, i.e. other coordinate choices (adapted to timelike observers) do not cause this. It is interesting to note that the same result was achieved in \cite{Swirl} when considering the Ehlers transformation on a Schwarzschild black hole, i.e. our metric when $B=e=0$.

\subsubsection[Dirac and Misner strings]{\large{Misner and Dirac strings}}

Dirac strings are, in a theoretical framework, the only way to explain the presence of magnetic monopoles and incorporate them in the Maxwell field equations. In particular, they can be seen as one-dimensional curves in spacetime which connect two oppositely charged magnetic ``Dirac monopoles'', as first described in Dirac's original 1931 article \cite{Dirac}. Their presence is implied whenever the following limits are not equal:

\begin{equation} \label{eq:dir1}
    \lim_{\theta\to 0} A_\phi = \frac{2 e^2 (12 B + B^3 e^2 + 8 e j)} {16 + 24 B^2 e^2 + B^4 e^4 + 32 B e^3 j + 16 e^4 j^2}
\end{equation}

\begin{equation} \label{eq:dir2}
\lim_{\theta\to \pi} A_\phi = \frac{2 e^2 (12 B + B^3 e^2 - 8 e j)} {16 + 24 B^2 e^2 + B^4 e^4 - 32 B e^3 j + 16 e^4 j^2}
\end{equation}

As we can clearly see, they do not coincide. What is nevertheless interesting to note is that, if we were to set the parameter $j=0$ and therefore be left with a magnetised Reissner-Nordström black hole, they would indeed coincide and Dirac strings would be absent. The same can be said if we set $e=0$ (as both limits would be zero), thereby having a magnetised, swirling Schwarzschild black hole. If we instead set $B=0$ (swirling Reissner-Nordström), the discrepancy remains: although this solution is only electrically charged, the rotation induces a magnetic dipole moment and $A_\phi\neq 0$.

A more interesting way to remove Dirac strings lies in a good choice of the rotation parameter $j$ (the same can be done with the electric charge): we set the difference between the two limits to zero and find the corresponding value of $j$, namely

\begin{equation} \label{j+-}
    j=\pm\frac{\sqrt{24B^2e^2+3B^4e^4-16}}{4e^2}
\end{equation}

These two values ensure that our spacetime is not plagued with Dirac strings.

Misner strings are the gravitational analogue of Dirac strings: wire-like singularities that are generally due to the presence of NUT charges instead of magnetic monopoles. Misner himself found these objects bizzarre and called them a ``counterexample to almost anything'' in \cite{Misner}. In order to visualise them we usually adopt Bonnor's interpretation, i.e. thinking of a Misner string as a semi-infinite, massless spinning rod (a source of angular momentum). For our purposes, a deeper knowledge of this topic is not necessary, but a thorough explanation may be found in \cite{Bonnor}

In order to check for their presence, we need to analyse the metric function $\omega_{el}(r,\theta)$, i.e. $\omega(r,\theta)$ as it appears in the electric LWP ansatz (\ref{eq:1.2}). It is found to be regular on the symmetry axis and asymptotically. We recognize its continuity through the following limits:

\[\lim_{\theta\to 0} \frac{g_{t\phi}}{g_{tt}}=\lim_{\theta\to\pi} \frac{g_{t\phi}}{g_{tt}}=0 \]

The regularity of $\omega_{el}(r,\theta)$ implies the absence of Misner strings.

\subsubsection[Conical singularities]{\large{Conical singularities}}

Axial conical singularities are a type of spacetime singularity visually similar to the pointed tip of a cone. In particular, they remove a hypothesis that is usually very important in General Relativity: the differentiable manifold ansatz. Mathematically, conical singularities manifest themselves as $\delta$-like divergences in the curvature (Riemann tensor), which is why they don't necessarily result as evident as other curvature singularities. As explained in \cite{Cmetric}, they can be interpreted as a ``cosmic string'', an infinitely long line under tension giving rise to a $\delta$-like contribution to the stress energy tensor $T_{\mu\nu}$.

Conical singularities can occur when a spin-spin interaction is present, because this generates a force without a consequent acceleration of the black hole. Since our spacetime rotates due to two separate contributions, we do have a kind of spin-spin interaction but no acceleration. It is then useful to check if our metric presents conical singularities.

Following the procedure found in \cite{Cmetric}, let us consider a small circle around the z-axis for $\theta=0$ and $\theta=\pi$ (while $t$ and $r$ are kept constant). If no angular defects are present, the ratio between the circle's circumference and its radius must be equal to $2\pi$. This is equivalent to the following condition\footnote{As usual, we consider $\phi\in[0,2\pi].$}:

\[ \lim_{\theta\to 0} \frac{1}{\theta} \int_{0}^{2\pi}\sqrt{\frac{g_{\phi\phi}}{g_{\theta\theta}}} \,d\phi = 2\pi = \lim_{\theta\to \pi} \frac{1}{\pi-\theta} \int_{0}^{2\pi}\sqrt{\frac{g_{\phi\phi}}{g_{\theta\theta}}} \,d\phi \]

Around the half axis $\theta=0$, we obtain:

\begin{equation} \label{eq:4.9}
    \lim_{\theta\to 0} \frac{1}{\theta} \int_{0}^{2\pi}\sqrt{\frac{g_{\phi\phi}}{g_{\theta\theta}}} \,d\phi = \frac{2\pi}{1 + \cfrac{3}{2}B^2e^2 + \cfrac{B^4 e^4}{16} + 2 B e^3 j + e^4 j^2}
\end{equation}

Which, unless we set some of our parameters to zero (either $e=0$ or $B=j=0$), is not exactly $2\pi$, which indicates the presence of a conical singularity.

The limit around the half axis $\theta=\pi$ yields:

\begin{equation} \label{eq:4.10}
    \lim_{\theta\to \pi} \frac{1}{\pi-\theta} \int_{0}^{2\pi}\sqrt{\frac{g_{\phi\phi}}{g_{\theta\theta}}} = \frac{2\pi}{1 + \cfrac{3}{2}B^2e^2 + \cfrac{B^4 e^4}{16} - 2 B e^3 j + e^4 j^2}  
\end{equation}

Once again, this is not exactly $2\pi$: this indicates a conical singularity of different \textit{conicity}, so not only are the limits not equal to $2\pi$, they don't even coincide. This is due to the facts that the denominator terms $2Be^3j$ have different signs, which means that the conical singularity originates from the coupling between $j$ and $B$ (and also $e$), which is a sort of spin-spin interaction. Consequently, if either $B$ or $j$ were equal to zero, leaving us with our two usual sub-metrics (swirling \textit{or} magnetised Reissner-Nordström), we would have two conical singularities of equal conicity.

There is a simple way to remove the deficit/excess angles (relative to $2\pi$) that we found, but we can only remove one of the two. Until now, we have thought of $\phi$ as an angular coordinate spanning the interval $[0,2\pi]$: what we can do now is redefine such coordinate as follows:

\begin{equation*}
    \phi'=\phi\cdot \Delta\phi
\end{equation*}

where $\Delta\phi$ is a constant to be chosen appropriately. This way, the new coordinate $\phi'$ spans a larger interval [$0,2\pi \Delta\phi$]. This rather trivial coordinate transformation changes the metric in the following way:

\begin{equation}
    ds^2=-\frac{f(r,\theta)}{|\Lambda(r,\theta)|^2}(dt-\omega'(r,\theta) d\phi\cdot\Delta\phi)^2+|\Lambda(r,\theta)|^2\left(Q(r)dt^2-\frac{dr^2}{Q(r)}-r^2d\theta^2 \right)
\end{equation}

(\ref{eq:4.9}) and (\ref{eq:4.10}) become:

\begin{equation} 
    \lim_{\theta\to 0} \frac{1}{\theta} \int_{0}^{2\pi}\sqrt{\frac{g_{\phi\phi}}{g_{\theta\theta}}} \,d\phi = \frac{2\pi\Delta\phi}{1 + \cfrac{3}{2}B^2e^2 + \cfrac{B^4 e^4}{16} + 2 B e^3 j + e^4 j^2}
\end{equation}

\begin{equation}
    \lim_{\theta\to \pi} \frac{1}{\pi-\theta} \int_{0}^{2\pi}\sqrt{\frac{g_{\phi\phi}}{g_{\theta\theta}}}\,d\phi = \frac{2\pi\Delta\phi}{1 + \cfrac{3}{2}B^2 e^2 + \cfrac{B^4 e^4}{16} - 2 B e^3 j + e^4 j^2}  
\end{equation}

In order to remove the conical singularity in $\theta=0$, $\Delta\phi$ is set to:

\begin{equation*}
    \Delta\phi=1 + \frac{3 B^2 e^2}{2} + \frac{B^4 e^4}{16} + 2 B e^3 j + e^4 j^2
\end{equation*}

We have now removed the conical singularity at $\theta=0$, but are left with the one at $\theta=\pi$, which as we mentioned earlier cannot be removed simultaneously. The following diagram should allow us to picture this situation more intuitively:

\begin{figure}[H]
    \centering
    \includegraphics[width=0.45 \textwidth]{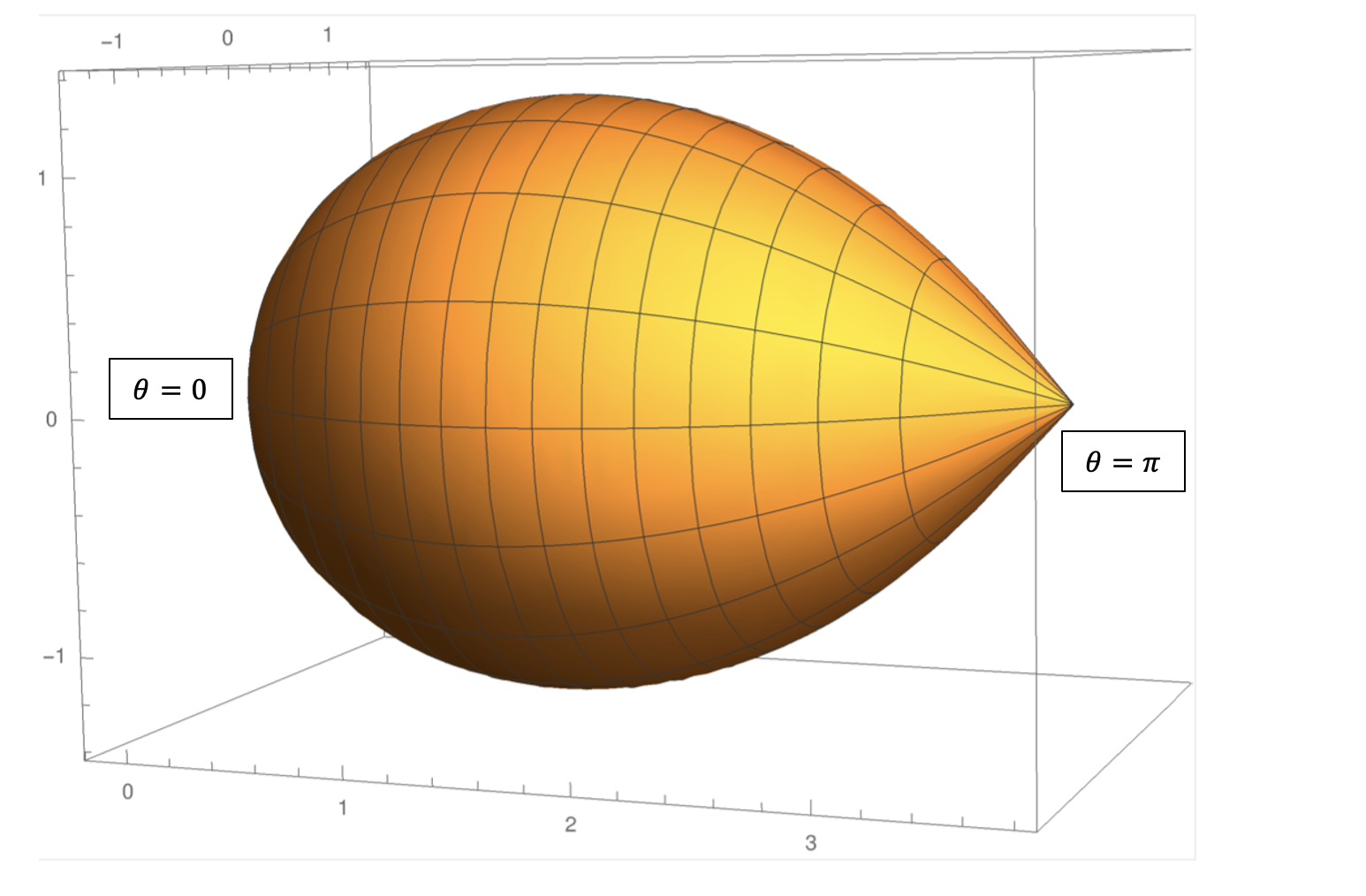}
    \caption{The surface of constant $t$, $r$ (embedded in the three-dimensional euclidean space $E^3$) we mentioned earlier is no longer a simple circle in our case: the round, regular shape at $\theta=0$ shows the successful removal of the first deficit angle, while the sharp edge at $\theta=\pi$ represents the remaining conical singularity.}
    \label{fig:Disegno2}
\end{figure}

As we mentioned, we are left with one non-removable conical singularity, which means that the spacetime cannot be regularised outside the event horizons. For this reason, in Section 5, we will attempt to remove the remaining nodal defect by introducing a new seed metric: the accelerated Reissner-Nordström black hole.

\subsection{Ergoregions}
Ergoregions are peculiar places that we can find in the neighbourhood of rotating black holes. Before exploring their presence and possible shapes in our own spacetime, we will introduce the concept using a simpler metric:

\begin{equation*} \begin{split}
    ds^2=&-\frac{\Delta-a^2\sin^2\theta}{\Sigma}dt^2-2a\sin^2\theta\frac{r^2+a^2-\Delta}{\Sigma}dtd\phi+\frac{(r^2+a^2)^2-\Delta a^2\sin^2\theta}{\Sigma}\sin^2\theta d\phi^2+\\&+\frac{\Sigma}{\Delta}dr^2+\Sigma d\theta^2 
\end{split} \end{equation*}

where

\begin{equation*} \begin{cases}
  \Delta=r^2-2mr+a^2 \\
  \Sigma=r^2+a^2\cos^2\theta
\end{cases} \end{equation*}

while $a=J/m$ is the angular momentum per unit mass. This is called the Kerr metric, a solution of the Einstein field equations in a vacuum found by Roy Kerr in 1963; it represents the simplest stationary black hole.

This metric presents two radii for which $g_{rr}=0$, which turn out to be event horizons. These are located at:

\begin{equation*}
    r_\pm=m\pm\sqrt{m^2-a^2}
\end{equation*}

Since the Kerr solution is not static but rather stationary, these are not ``Killing horizons'' for the Killing vector $\partial_t$, i.e. $g_{tt}$ is not equal to zero at $r_\pm$.

Instead, $g_{tt}$ reaches zero at a surface that lies outside the outer event horizon $r_+$:

\begin{equation*}
    r(\theta)=m+\sqrt{m^2-a^2\cos^2\theta}
\end{equation*}

This is called the stationary limit surface, where the black hole's frame dragging becomes so extreme that it is impossible for an observer to remain static relative to distant stars. In fact, all observers at and inside this surface must orbit the black hole with positive angular velocity (i.e. in the black hole's direction of rotation).

Let us represent these neighbouring regions outside the Kerr black hole with the following drawing:

\begin{figure}[H]
    \centering
    \includegraphics[width=0.45 \textwidth]{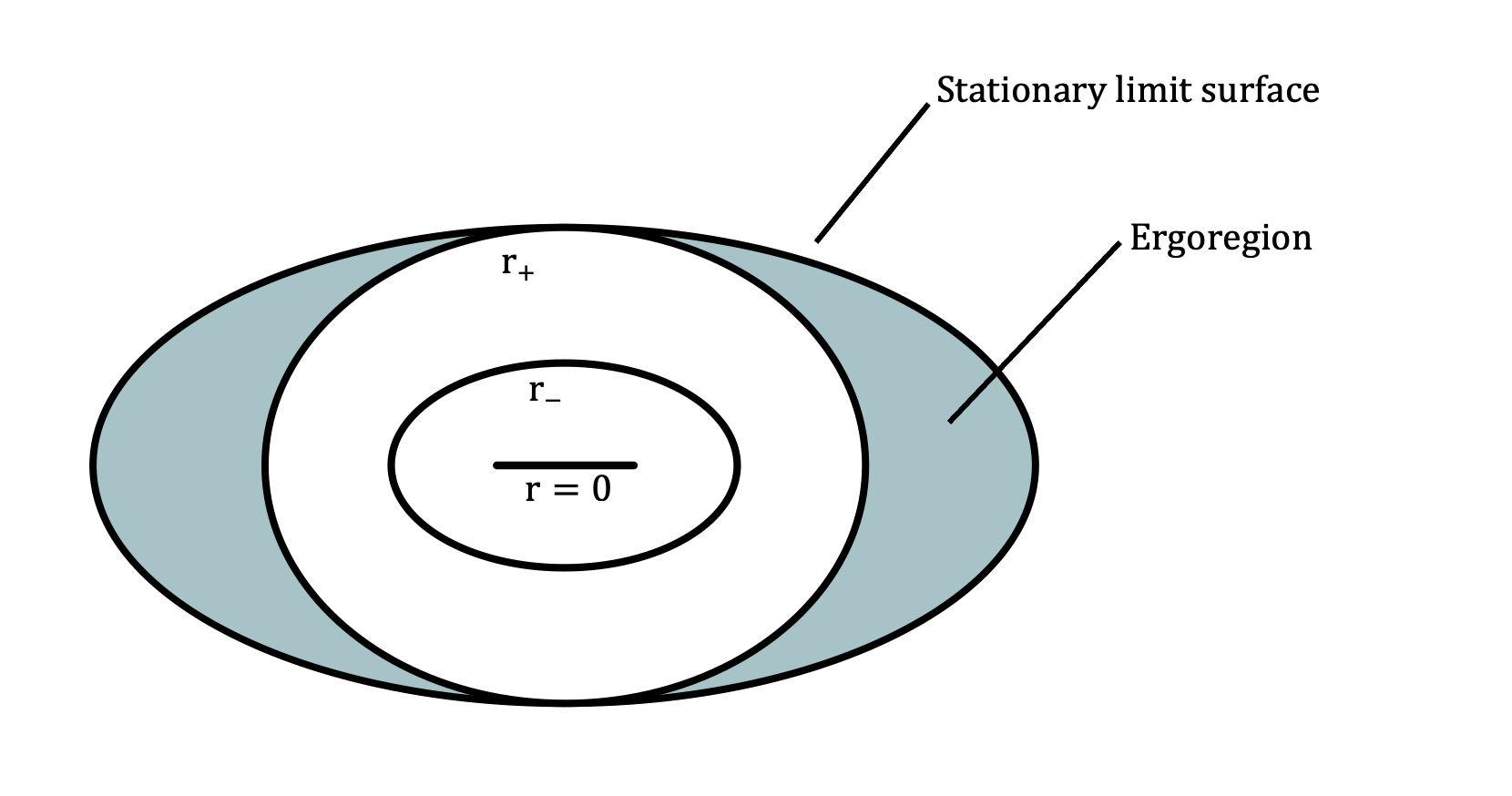}
    \caption{The various surfaces surrounding the rotating Kerr black hole: the inner and outer event horizons and the stationary limit surface. The latter two surfaces are connected by a region called ``ergosphere'' or ergoregion.}
    \label{fig:Disegno3}
\end{figure}

As we can see in the above representation, there is a region between the outer event horizon and the stationary limit surface: the  ergoregion. Since it lies inside the stationary limit surface, only stationary observers with positive angular velocity can exist within this region, however nothing forbids them from moving closer to or further from $r_+$, which means that they are free to leave the ergoregion.

A very interesting feature of ergoregions is the Penrose process, which exploits the fact that a particle with negative energy could theoretically exist in such a region, due to energy being defined as $E=-K^\mu p_\mu$ and the Killing vector $K^\mu$ becoming spacelike inside the ergosphere, while the four-momentum remains timelike. This represents a way to extract energy from a black hole.
\newline
The in-depth explanation of this process is not necessary for the purposes of this thesis, but extremely good ones can be found in \cite{Gravitation} and \cite{Carroll}.

Through an analysis of our metric's $g_{tt}$ component's regions of positivity, we will find that our metric possesses ergoregions, just like the Kerr, the magnetised Reissner-Nordström (\cite{Gibb}) and the swirling Schwarzschild (\cite{Swirl}) solutions. Since our metric was obtained using not one but two transformations, it seems appropriate to discuss the cases $j=0$ (magnetised Reissner-Nordström), $B=0$ (swirling Reissner-Nordström) and $j,B\neq 0$ (the full metric) separately. 

\subsubsection[The magnetised Reissner-Nordström case]{\large{The magnetised Reissner-Nordström case}}

Although the magnetised Reissner-Nordström metric's ergoregions were already found by Gibbons, Mujtaba and Pope in \cite{Gibb}, we want to analyse them using more parameter choices.

Our metric component is

\begin{equation} \label{eq:gtt}
    g_{tt}=-\frac{f(r,\theta)}{|\Lambda(r,\theta)|^2}\omega'(r,\theta)^2-|\Lambda(r,\theta)|^2Q(r)
\end{equation}

In this case, the function $\omega'(r,\theta)$ can be written as

\begin{equation*}
    \omega'(r,\theta)=2e\frac{B}{r}+\frac{1}{2} eB^3r(1+Q(r)\cos^2\theta)
\end{equation*}

Therefore, by simply glancing at (\ref{eq:gtt}), we can see that the second term vanishes on the outer horizon $r_+=r+\sqrt{m^2-e^2}$ because of $Q(r)$, while the first term gives a positive contribution wherever $f(r,\theta)\neq 0$, i.e. where $\sin^2\theta\neq 0$ since we are considering $r$ nonzero. This means that $g_{tt}$ will be positive right outside the outer horizon, just like in the Kerr case.

In order to best visualize the complete ergoregion, it is best to use cylindrical coordinates

\begin{equation*} \begin{cases}
    \rho=r\sin\theta \\
    z=r\cos\theta
\end{cases} \end{equation*}

as done in \cite{Swirl} and \cite{Gibb}. While the complete expression of $g_{tt}$ in these coordinates is quite long and unnecessary for our purposes, let us instead report its expansion for large z:

\begin{equation} \label{eq:gj}
    g_{tt}(\rho,z)\approx\frac{16B^6e^2(z^2-2mz)\rho^2}{16+B^4(e^2+\rho^2)^2+8B^2(3e^2+\rho^2)}
\end{equation}

Clearly, if we were to hold $\rho$ fixed in the large z limit, $g_{tt}$ would grow arbitrarily large (and positive).

Therefore, the ergoregions are not only located in $r_+$'s proximity but also close to the z axis, where they extend to infinity.

We will now proceed to graphically visualise these regions for different choices of our parameters $e$ and $B$. Let us note that all graphs regarding ergoregions presented from now on will represent a cross-section through the ergoregion, while the full ergoregion is the surface of revolution that can be obtained by rotating the cross-section around the vertical axis.  

First, we fix the mass ($m=1$) and intrinsic electric charge ($e=0.5$) while varying the magnetic field of the background $B$.

\begin{figure}[H]
    \centering
    \includegraphics[width=0.45 \textwidth]{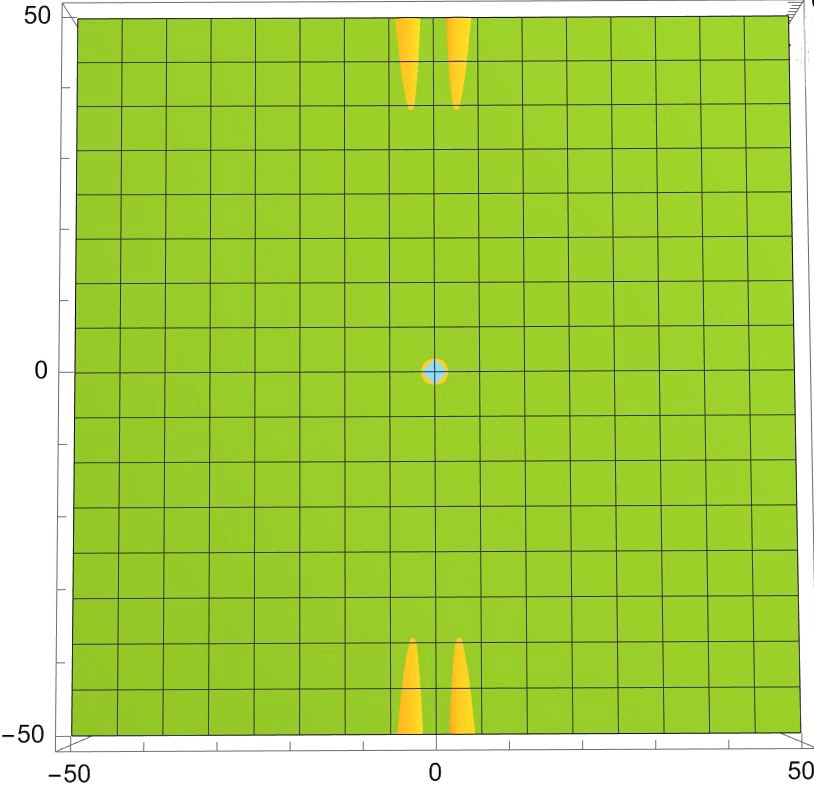}
    \caption{Ergoregion cross section for a Reissner-Nordström black hole embedded in a magnetic universe, with parameters $m=1$, $B=0.6$ and $e=0.5$. A magnetic field parameter below $B=1$ yields three main regions that are very far apart from one another, so much so that the graph's field of view had to be increased to see them all. In particular, we can see that the ergoregions extend to infinity in the positive and negative z directions, as for all graphs pertaining to this and the next subcase.  }
    \label{fig:Disegno4b}
\end{figure}

\begin{figure}[H]
    \centering
    \includegraphics[width=0.45 \textwidth]{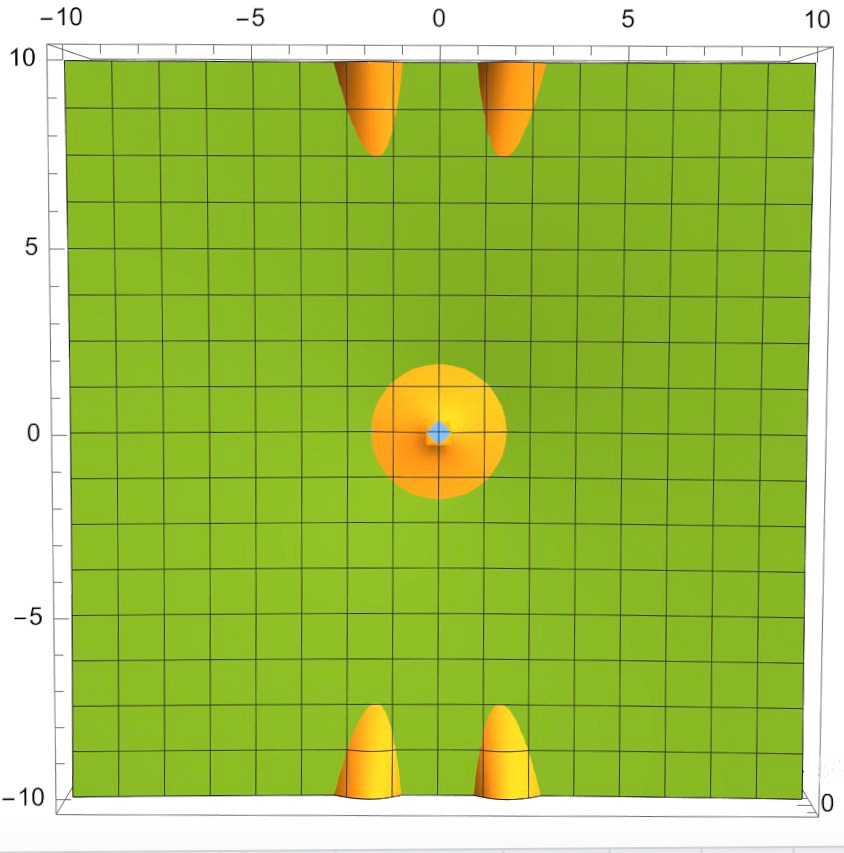}
    \caption{Ergoregion cross section of a Reissner-Nordström black hole embedded in a magnetic universe, with parameters $m=1$, $B=1$ and $e=0.5$. Compared to the previous graph, the three main regions are closer. As in all ergoregion graphs presented, a blue spherical hole is visible at the center of the central yellow region: this represents the black hole's outer event horizon. }
    \label{fig:Disegno5b}
\end{figure}

\begin{figure}[H]
    \centering
    \includegraphics[width=0.45 \textwidth]{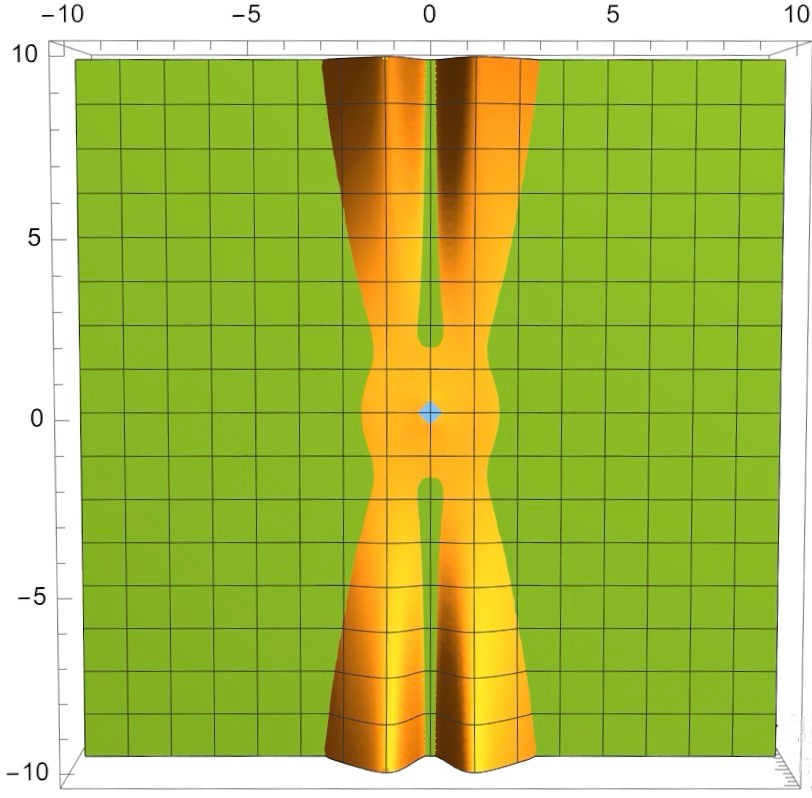}
    \caption{Ergoregion cross section of a Reissner-Nordström black hole embedded in a magnetic universe, with parameters $m=1$, $B=2$ and $e=0.5$. For rising values of $B$ the three main regions become closer and closer until they are attached, as can be seen in this case. This graph qualitatively reproduces the results found in \cite{Gibb}.}
    \label{fig:Disegno6b}
\end{figure}

\vspace{0.2cm}

Let us now analyse the $e=0.75$ case to see how the ergoregion cross sections change accorging to the intrinsic electric charge's value.

\vspace{1cm}

\begin{figure}[H]
    \centering
    \includegraphics[width=0.45 \textwidth]{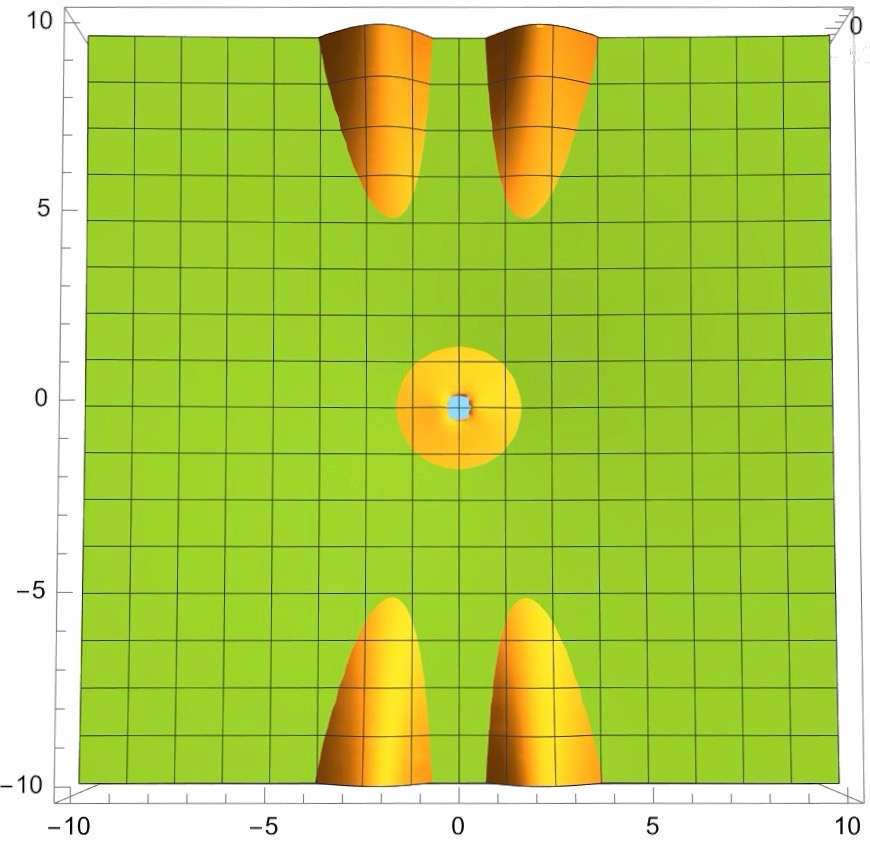}
    \caption{Ergoregion cross section of a Reissner-Nordström black hole embedded in a magnetic universe, with parameters $m=1$, $B=1$ and $e=0.75$. Compared to Figure \ref{fig:Disegno5b}, the main difference is that the upper and lower main regions come closer to the central one in this case. In addition, the central region becomes slightly smaller. The $B=0.6$ case presents the same modifications.}
    \label{fig:Disegno7b}
\end{figure}

\begin{figure}[H]
    \centering
    \includegraphics[width=0.45 \textwidth]{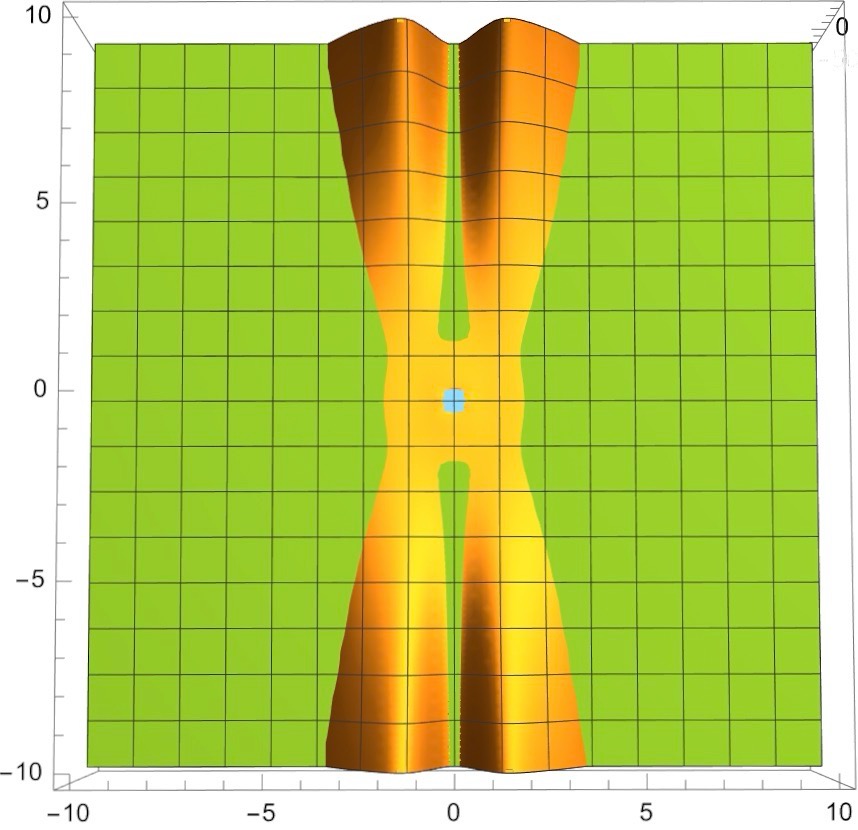}
    \caption{Ergoregion cross section of a Reissner-Nordström black hole embedded in a magnetic universe, with parameters $m=1$, $B=2$ and $e=0.75$. Similarly to the $B=1$ case, the central region is smaller than it was in Figure \ref{fig:Disegno6b}. Instead, the upper and lower regions appear wider, reflecting the fact that they ``approached'' to the central region as they did in Figure \ref{fig:Disegno7b}. }
    \label{fig:Disegno8b}
\end{figure}

\subsubsection[The swirling Reissner-Nordström case]{\large{The swirling Reissner-Nordström case}}

In this second sub-case, we turn off the external magnetic background. If we were to also set $e=0$ we would have the swirling Schwarzschild metric, whose ergoregions are analysed in \cite{Swirl}.

The general expression for $g_{tt}$ is clearly still (\ref{eq:gtt}), while the metric function $\omega'(r,\theta)$ becomes

\begin{equation*}
        \omega'(r,\theta)=\frac{4j(e^2+r(r-2m))}{r}\cos\theta
\end{equation*}

Once again, (\ref{eq:gtt})'s second term vanishes on the outer horizon $r_+$ and the first one remains positive wherever $\sin^2\theta\neq 0$. Hence, $g_{tt}$ will be positive right outside the outer horizon, just like in the previous subcase.

We switch to cylindrical coordinates and report $g_{tt}$'s approximate expression for large z:

\begin{equation} \label{eq:gB}
    g_{tt}(\rho,z)\approx\frac{16j^2(z^2-4mz)\rho^2}{1+j^2(e^2+\rho^2)^2}
\end{equation}

We see that, analogously to the previous subcase, $g_{tt}$ can become arbitrarily large as $z$'s value increases and $\rho$'s remains fixed.

Hence, the ergoregion extends along the z axis in addition to being located in the outer proximity of $r_+$. We can see this graphically for some different values of our parameters $e$ and $j$.

We start by fixing the mass ($m=1$) and intrinsic electric charge ($e=0.5$) while varying the angular velocity of the background $j$.

\hspace{0.2cm}

\begin{figure}[H]
    \centering
    \includegraphics[width=0.45 \textwidth]{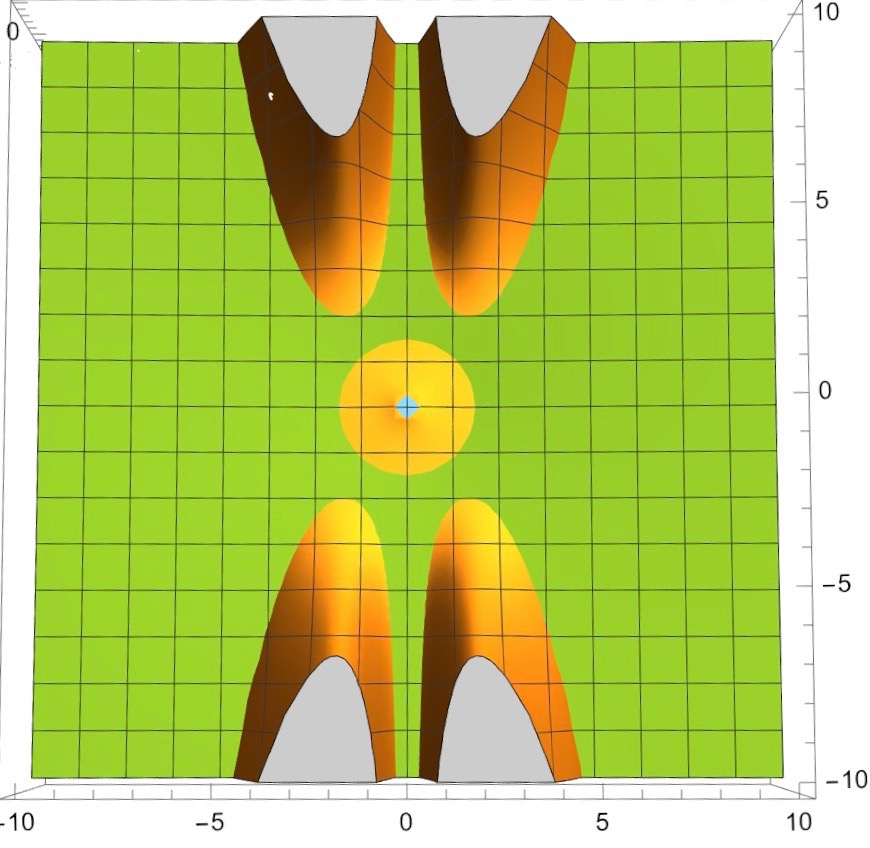}
    \caption{Ergoregion cross section of a Reissner-Nordström black hole embedded in a rotating universe, with parameters $m=1$, $e=0.5$ and $j=0.3$. The ergoregion is present in the outer proximity of the outer event horizon and it extends to infinity on the z axis, in positive and negative directions. The two upper and the two lower yellow regions are not connected to the central one, however they are connected to each other (respectively) at larger distances. Let us note that the grey regions still constitute an ergoregion but $g_{tt}$'s highly positive value exceeds the field of view (from a depth perspective) and they are therefore not yellow.  }
    \label{fig:Disegno4a}
\end{figure}

 \begin{figure}[H]
    \centering
    \includegraphics[width=0.45 \textwidth]{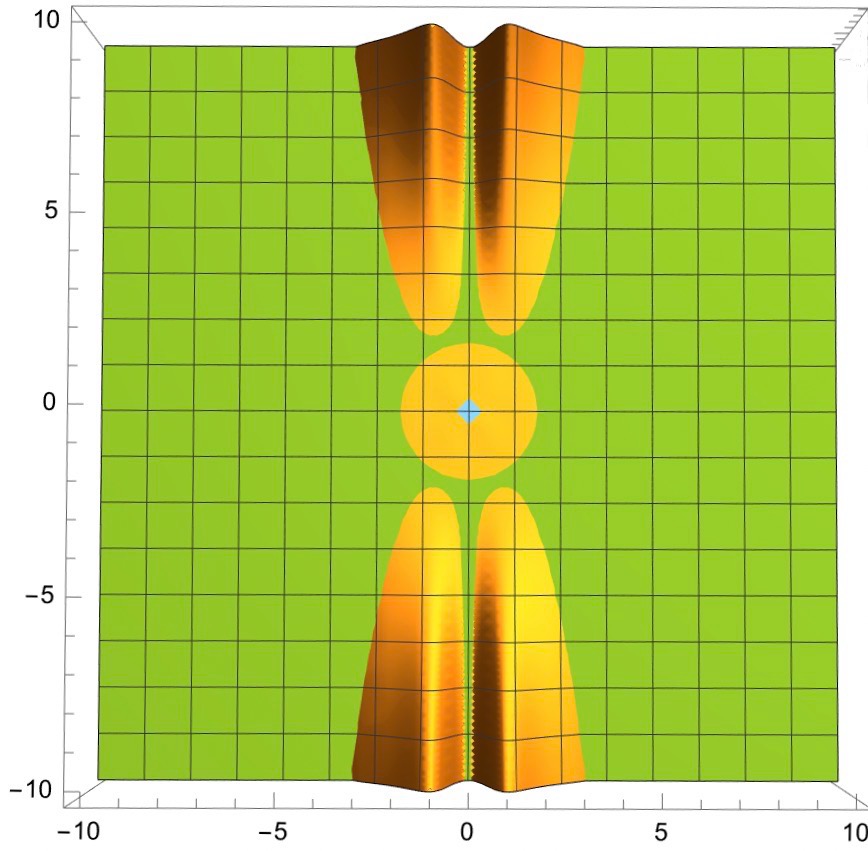}
    \caption{Ergoregion cross section of a Reissner-Nordström black hole embedded in a rotating universe, with parameters $m=j=1$ and $e=0.5$. The yellow regions are very similar to the previous ($j=0.3$) case: the only visible difference is that, in this case, the upper and lower ``legs'' are narrower and $g_{tt}$'s value no longer exceeds the field of view. }
    \label{fig:Disegno5a}
\end{figure}

\begin{figure}[H]
    \centering
    \includegraphics[width=0.45 \textwidth]{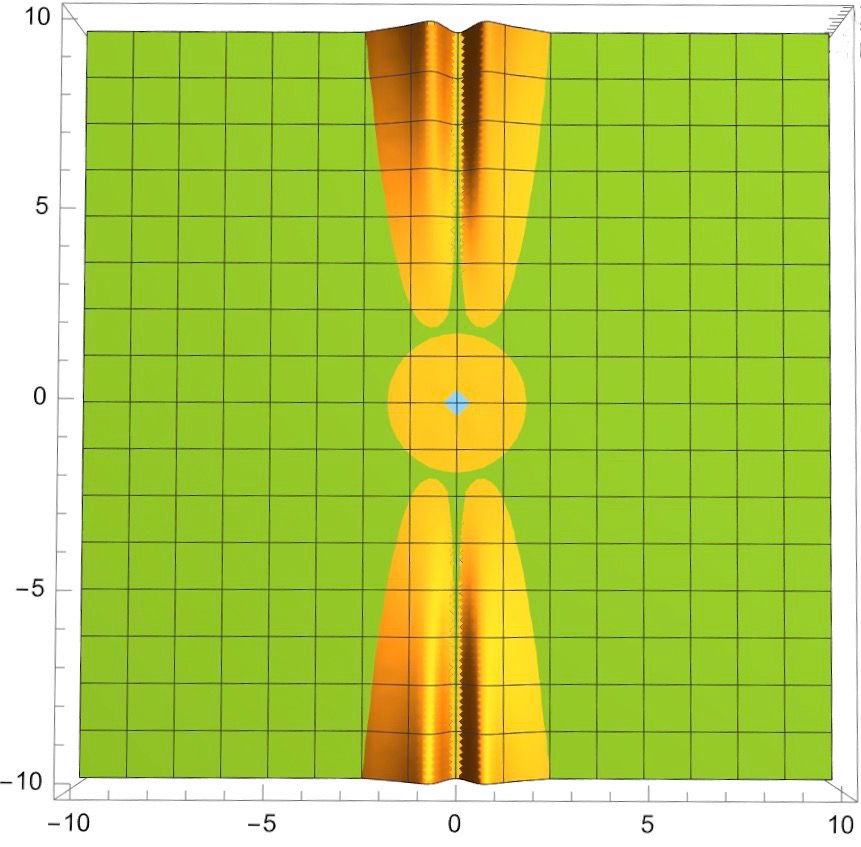}
    \caption{Ergoregion cross section of a Reissner-Nordström black hole embedded in a rotating universe, with parameters $m=1$, $j=2$ and $e=0.5$. Compared to the previous graph, this presents narrower ``legs'' that are closer together, hence they become attached at smaller distances. }
    \label{fig:Disegno5a}
\end{figure}

\vspace{0.2cm}

Let us now take a different value for the electric charge, namely $e=0.75$, to see how it influences the ergoregion's shape in the $j=0.3$ case (the changes are analogous in the $j=1$ case).

\vspace{1cm}

 \begin{figure}[H]
    \centering
    \includegraphics[width=0.45 \textwidth]{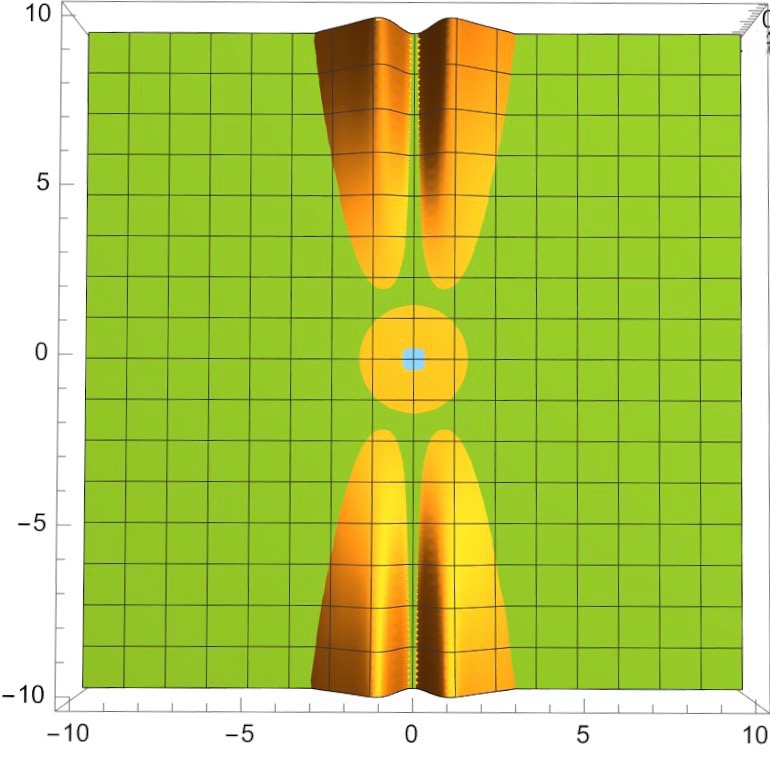}
    \caption{Ergoregions (cross section) for a Reissner-Nordström black hole embedded in a rotating universe, with parameters $m=j=1$ and $e=0.75$. If compared to Figure \ref{fig:Disegno4a}, the main difference is the size of the central region, which is smaller in this case. }
    \label{fig:Disegno6a}
\end{figure}

\vspace{1cm}

\subsubsection[The swirling, magnetic Reissner-Nordström case]{\large{The swirling, magnetic Reissner-Nordström case}}

We finally analyse the complete metric's ergoregions, for which $g_{tt}$'s general expression (\ref{eq:gtt}) still holds. Let us report $\omega'(r,\theta)$'s expression for completeness:

\begin{equation*}
        \omega'(r,\theta)=\frac{B^3er^2-4Be+\cos\theta(e^2+r(r-2m))(B^3e \cos\theta-8j)}{2r}
\end{equation*}

The same argument used in the two previous subcases is applicable here, so, once again, we have a region of positivity for $g_{tt}$ just outside the outer horizon $r_+$. Applying cylindrical coordinates and taking the $z\xrightarrow{}\infty$ limit, we obtain

\begin{equation} \label{eq:gjB}
    g_{tt}(\rho,z)\approx\frac{16(B^3e-4j)z(4j(z-4m)+B^3e(z-2m))\rho^2}{16+32Bej(e^2+\rho^2)+B^4(e^2+\rho^2)^2+16j^2(e^2+\rho^2)^2+8B^2(3e^2+\rho^2)}
\end{equation}

Naturally, since this case has nonzero $j$ and $B$, the expression is more complicated than the ones we saw earlier: $g_{tt}$'s trend is not as straightforward at first glance. Let us then look at the coefficient corresponding to the highest power of $z$, i.e. $z^2$:

\begin{equation*}
    16(B^3e-4j)^2\rho^2z^2
\end{equation*}

We do not necessarily have an arbitrarily large value for $g_{tt}$ in this case: it depends on the parameters. In particular, if

\begin{equation*}
    j=\frac{B^3e}{4}
\end{equation*}

then $g_{tt}$ is exactly zero. Therefore, in order to have $g_{tt}$'s arbitrary growth for large $z$ and fixed $\rho$, we need

\begin{equation*}
    j\neq\frac{B^3e}{4}
\end{equation*}

To see this graphically, we fix the mass ($m=1$) and intrinsic electric charge ($e=0.5$) while varying the angular velocity of the background $j$ and the magnetic field of the background $B$.

It is important to note that now, unlike seen in the two previous cases, the ergoregions will not be symmetrical with respect to the vertical axis (z). The reason for this is simple: the Lorentz force causes a uniform rotation, in the same direction at any $r, \theta, \phi$. The swirling background instead causes a whirlpool rotation: if one thinks of a sphere immersed in such a universe, the spacetime around the north and south poles would rotate in opposite directions. A background that is both magnetic and swirling would therefore cause a highly non-uniform rotation\footnote{For instance, the frame dragging can decrease in some points in spacetime and increase in others.} of spacetime (\ref{eq:4.8}), which reflects on the ergoregions' symmetry.

\begin{figure}[H]
    \centering
    \includegraphics[width=0.45 \textwidth]{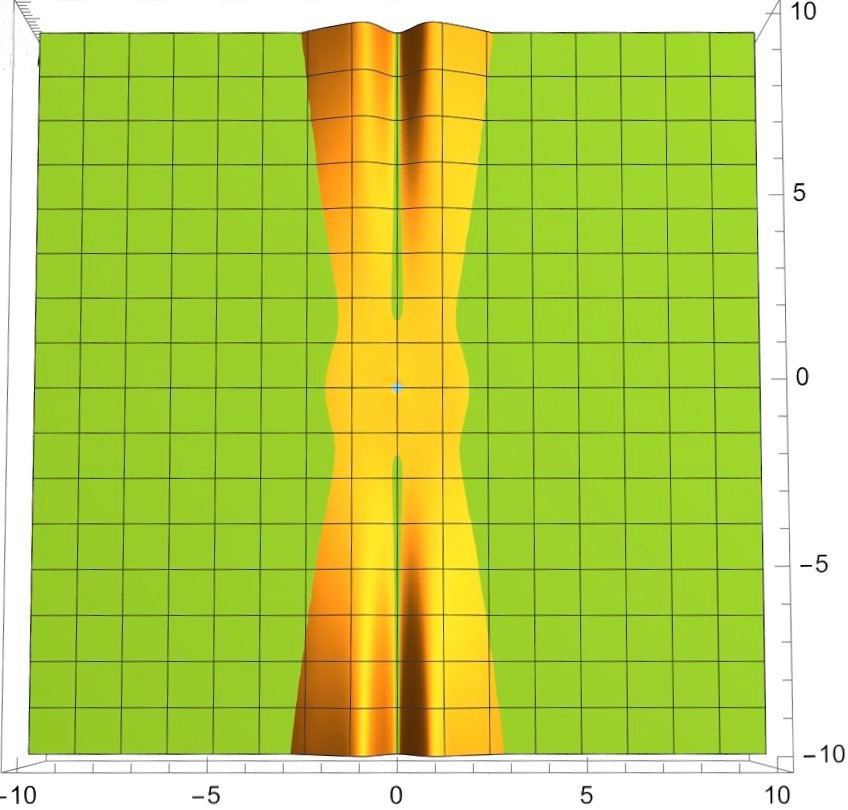}
    \caption{Ergoregion cross section of a Reissner-Nordström black hole embedded in a rotating, swirling universe with parameters $m=1$, $e=0.5$, $j=0.3$ and $B=3$. This graph does not present extremely evident asymmetries because $j$'s value is much smaller than $B$'s, hence the coupling of the two rotations is almost negligible. The contribution of the magnetic background is dominant, which explains why this graph looks very similar to Figure \ref{fig:Disegno6b}. Once again, the ergoregions extend to infinity in positive and negative z directions.  }
    \label{fig:Disegno4ab}
\end{figure}

\begin{figure}[H]
    \centering
    \includegraphics[width=0.45 \textwidth]{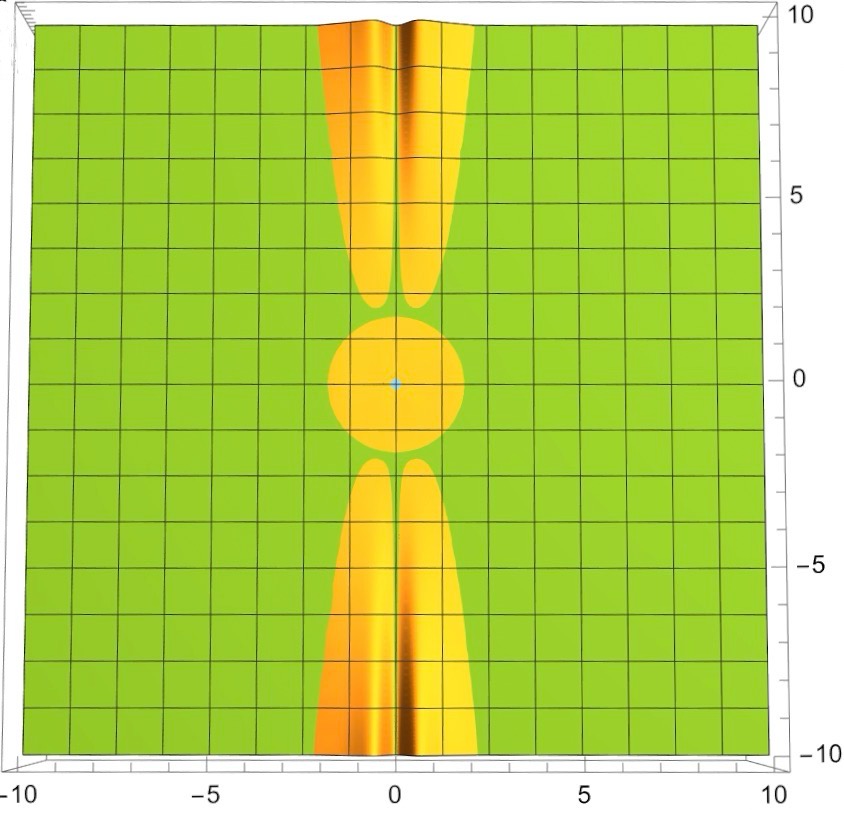}
    \caption{Ergoregion cross section of a Reissner-Nordström black hole embedded in a rotating, swirling universe with parameters $m=1$, $e=0.5$, $j=3$ and $B=0.3$ (the opposite of the previous graph). Once again, asymmetries are not evident because $j$ and $B$ have very different values. In this case, the contribution of the swirling background is dominant, explaining this graph's similarity to Figure \ref{fig:Disegno5a}.  }
    \label{fig:Disegno5ab}
\end{figure}

\begin{figure}[H]
    \centering
    \includegraphics[width=0.45 \textwidth]{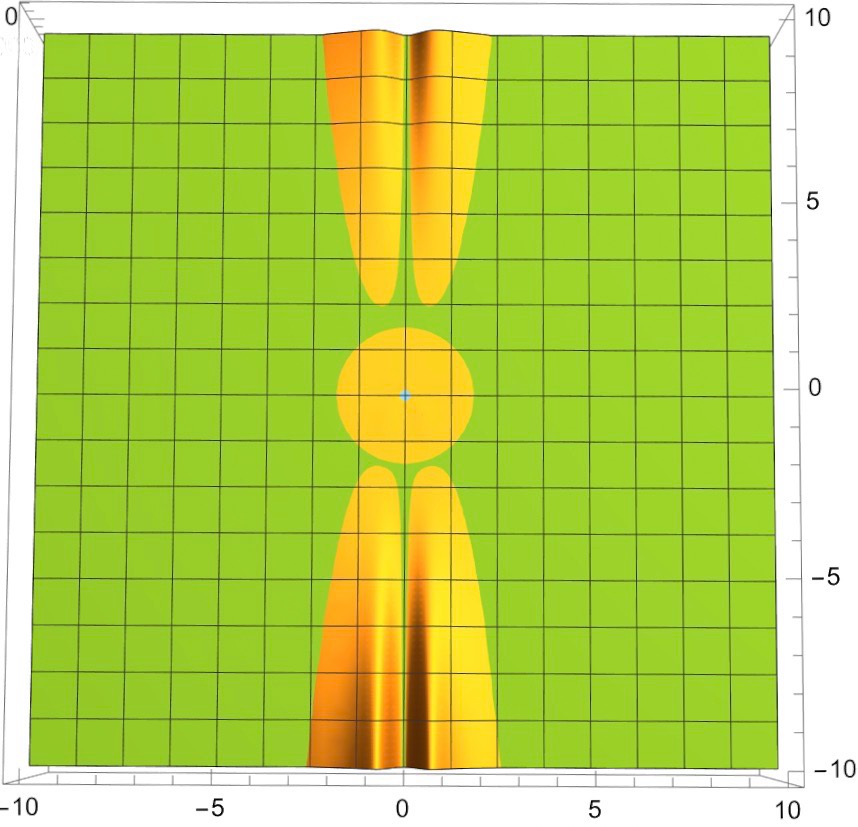}
    \caption{Ergoregion cross section of a Reissner-Nordström black hole embedded in a rotating, swirling universe with parameters $m=1$, $e=0.5$, $j=2$ and $B=1$. $j$'s value is now twice $B$'s: the two contributions to the rotation can be considered more comparable, which means that asymmetries will be evident. For instance, it can be clearly seen that the lower ``legs'' are wider and closer to the central region compared to the upper ``legs''.   }
    \label{fig:Disegno6ab}
\end{figure}

\begin{figure}[H]
    \centering
    \includegraphics[width=0.45 \textwidth]{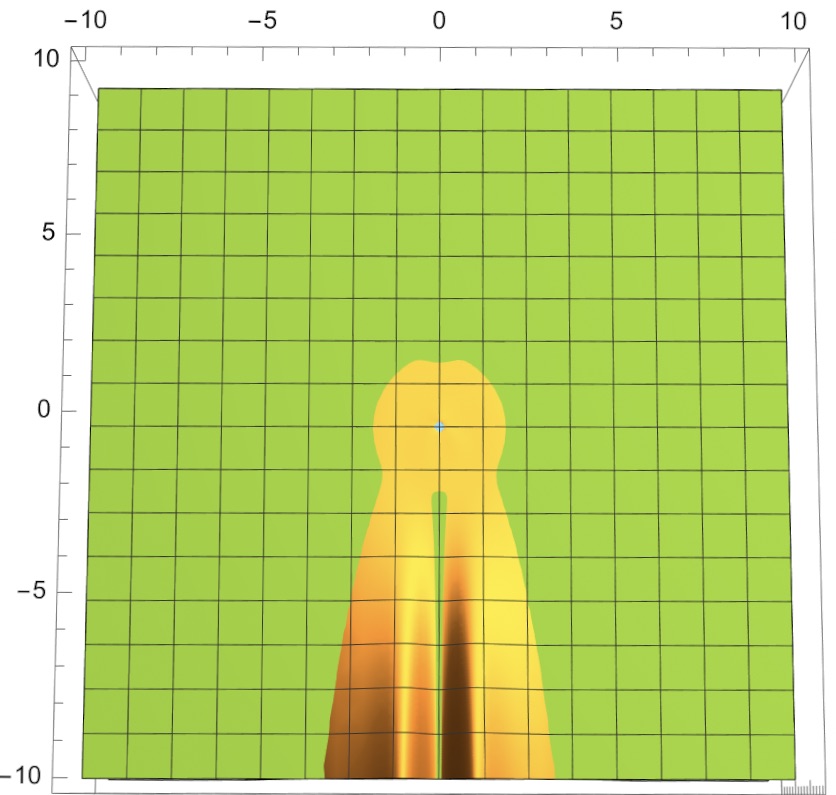}
    \caption{Ergoregion cross section of a Reissner-Nordström black hole embedded in a rotating, swirling universe with parameters $m=1$, $e=0.5$, $j=1$ and $B=2$. $B$'s value is now twice $j$'s, so evident asymmetries are to be expected, like in the previous graph. In this case, the two upper ``legs'' have disappeared at very large distances while the lower two are attached to the central region.  }
    \label{fig:Disegno7ab}
\end{figure}

\begin{figure}[H]
    \centering
    \includegraphics[width=0.45 \textwidth]{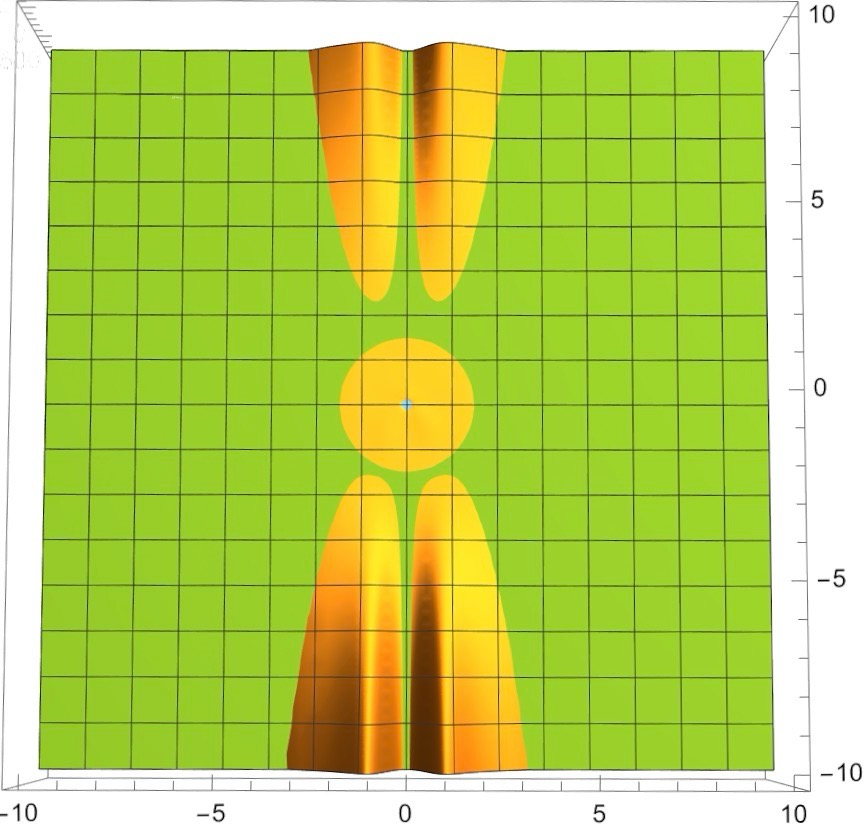}
    \caption{Ergoregions (cross section) for a Reissner-Nordström black hole embedded in a rotating, swirling universe with parameters $m=1$, $e=0.5$, $j=1$ and $B=1$. $j$ and $B$'s contributions are now equally influential and the most evident asymmetry is precisely the one we previously found in Figure \ref{fig:Disegno6ab}: the lower ``legs'' are wider and closer to the central region compared to the upper ``legs''.  }
    \label{fig:Disegno8ab}
\end{figure}

\clearpage

\section{The accelerated solution}
In this chapter, we will derive a generalisation of the solution we found, namely the accelerated version. We will follow the procedure used in chapter 3, taking the accelerated Reissner-Nordström metric as a seed. The latter, also known as a charged C-metric, can be regarded as a black hole that accelerates because of the presence of cosmic strings (or struts), which mathematically corresponds to a conical singularity. More information on C-metrics and their general properties can be found in Appendix A.

\subsection{Accelerated Reissner-Nordström black hole}
The generalised version of the Reissner-Nordström spacetime featuring an acceleration can be expressed as follows in spherical coordinates, as done in \cite{Paircr}:

\begin{equation} \label{eq:5.0}
    ds^2=\frac{1}{(1+Ar\cos\theta)^2}\left(-Q(r)dt^2+\frac{dr^2}{Q(r)}+\frac{r^2d\theta^2}{P(\theta)}+r^2\sin^2\theta P(\theta) d\phi^2\right)
\end{equation}

where

\begin{equation*}
    Q(r)=(1-A^2r^2)\left(1-\frac{2m}{r}+\frac{e^2}{r^2}\right)
\end{equation*}

\begin{equation*}
    P(\theta)=1+2mA\cos\theta+A^2e^2\cos^2\theta
\end{equation*}

The constant A represents the acceleration parameter, while $m$ is the mass and $e$ is the electric charge just like in (\ref{eq:3.0}). This metric is compatible with the same potential as (\ref{eq:3.0}).

As we mentioned, metric (\ref{eq:5.0}) is a charged C-metric: a non-asymptotically flat generalisation of the Reissner-Nordström spacetime.

\subsubsection[Comparison with magnetic LWP]{\large{Comparison with magnetic LWP}}
We will now compare (\ref{eq:5.0}) to the magnetic LWP metric and follow the procedure of chapter 3. To make this discussion clearer, let us recall the expression of the magnetic LWP metric:

\begin{equation*}
    ds^2=-f(d\phi-\omega(r,\theta) dt)^2+\frac{1}{f}(\rho^2d\phi^2-e^{2\gamma}(d\rho^2+dz^2))
\end{equation*}

The coordinate change required to connect spherical and cylindrical Weyl coordinates is the following:

\begin{equation} \label{eq:5.1}
\begin{cases}
\rho(r,\theta)=\cfrac{r\sin\theta\sqrt{Q(r)P(x)}}{(1+Ar\cos\theta)^2} \\

z(r,\theta)=\cfrac{Ar\cos\theta(r+m(Ar\cos\theta-1)-Ae^2\cos\theta)}{(1+Ar\cos\theta)^2}
\end{cases}
\end{equation}

Once again, we found $\rho(r,\theta)$ by simply comparing the two metrics, while $z(r,\theta)$ involves solving a differential equation. These expressions are more complicated than the ones found in chapter 3, however the limit $A\xrightarrow{}0$ yields exactly (\ref{eq:3.1}).

In particular, we must note that all the functions we will find throughout this chapter do \textit{not} directly simplify to those of the non-accelerating case if we simply put $A=0$, due to $A$'s position in the metric's structure. This is indeed a general property of accelerated metrics and the issue can be easily bypassed by taking the $A\xrightarrow{}0$ limit.

Our seed metric, though accelerating, remains non rotating, so $\omega(r,\theta)=0$. The juxtaposition of the $d\phi^2$ and $d\rho^2+dz^2$ terms yields:

\begin{equation} \begin{cases}
f(r,\theta)=-\cfrac{r^2\sin^2\theta P(\theta)}{(1+Ar\cos\theta)^2} \\

e^{2\gamma(r,\theta)}=\cfrac{r^4\sin^2\theta P(\theta)}{N(r,\theta)(1+Ar\cos\theta)}
\end{cases} 
\end{equation}

where

\begin{equation*} \begin{split}
    N(r,\theta)=&(A (A(2mr- e^2)+\cos\theta(r-2m) + A^2 e^2 r\cos\theta-1)) (2 m (r - 
      A e^2 \cos\theta) \\&(A r \cos\theta-1) + (1 + A^2 e^2) (r^2 + e^2 \cos^2\theta) + 
   m^2 (\sin^2\theta(1-A^2r^2)-4Ar\cos\theta)
\end{split} \end{equation*}

In the $A\xrightarrow{}0$ limit, these change precisely into (\ref{eq:3.4}).

We must now look for the Ernst potentials $\mathbfcal{E}$ and $\boldsymbol{\Phi}$: we begin by using equation (\ref{eq:3.6}) to find $\tilde{A_t}$. Having a new metric and a new coordinate transformation, the gradient takes on a different form:

\begin{equation}
    \nabla\propto\sqrt{\Delta(r)}\partial_r\hat{r} +\sqrt{P(\theta)}\partial_\phi\hat{\phi}
\end{equation}

We still have the regular Reissner-Nordström solution's four-potential with $A_t=-\cfrac{e}{r}$ \hspace{0.05cm} and $A_\phi=0$, so (\ref{eq:3.6}) gives us $\tilde{A_t}=-ecos\theta$. Therefore

\begin{equation} \label{eq:5.3}
\boldsymbol{\Phi}(\theta)=-ie\cos\theta
\end{equation}

Now looking for $\mathbfcal{E}$, we use equation (\ref{eq:3.8}) to find h. With the same arguments used in chapter 3, our definition $h(r,\theta)=0$ is justified. This gives us

\begin{equation} \label{eq:5.4}
\mathbfcal{E}(r,\theta)=-\frac{r^2\sin^2\theta P(\theta)}{(1+Ar\cos\theta)^2}-e^2\cos^2\theta
\end{equation}

\subsection{Finding the new, accelerated solution}
We have found all of the necessary functions related to our seed metric, which means that we can now apply the hybrid Harrison-Ehlers transformation, which we report for convenience:

\begin{equation*} \begin{cases}
  \boldsymbol{\Phi}'=\cfrac{\boldsymbol{\Phi}-\cfrac{B}{2}\mathbfcal{E}}{1+ij\mathbfcal{E}+B\boldsymbol{\Phi}-\cfrac{B^2}{4}\mathbfcal{E}} \\
 \mathbfcal{E}'=\cfrac{\mathbfcal{E}}{1+ij\mathbfcal{E}+B\boldsymbol{\Phi}-\cfrac{B^2}{4}\mathbfcal{E}}
\end{cases} \end{equation*}

By merely using $\mathbfcal{E}$ and $\boldsymbol{\Phi}$'s definitions and expressions, we obtain:

\begin{equation} \label{eq:5.5}
    A_\phi'(r,\theta)=\frac{1}{|\Lambda(r,\theta)|^2}\left(e^2B\cos^2\theta-\frac{B}{2}\mathbfcal{E}\left(1-\frac{B^2}{4}\mathbfcal{E} \right)-je\mathbfcal{E} \cos\theta \right)
\end{equation}

\begin{equation} \label{eq:5.6}
    \tilde{A_t}'(r,\theta)=\frac{1}{|\Lambda(r,\theta)|^2}\left(j\frac{B}{2}\mathbfcal{E}^2-e\cos\theta\left(1+\frac{B^2}{4}\mathbfcal{E}\right)\right)
\end{equation}

\begin{equation} \label{eq:5.7}
    h'(r,\theta)=\frac{Be\cos\theta\mathbfcal{E}-j\mathbfcal{E}^2}{|\Lambda(r,\theta)|^2}
\end{equation}

\begin{equation} \label{eq:5.8}
    f'(r,\theta)=\frac{f(r,\theta)}{|\Lambda(r,\theta)|^2}
\end{equation}

where, in analogy with equations (\ref{eq:4.2}) through (\ref{eq:4.5}), we defined $\Lambda(r,\theta)$ as the transformation's denominator:

\begin{equation*}
    \Lambda(r,\theta)=1+ij\mathbfcal{E}+B\boldsymbol{\Phi}-\cfrac{B^2}{4}\mathbfcal{E}
\end{equation*}

Again, using (\ref{eq:2.18}) and (\ref{eq:2.19}) we find that $\gamma'(r,\theta)=\gamma(r,\theta)$, so the only  functions yet to be determined are $\omega'(r,\theta)$ and $A_t'(r,\theta)$. Using, respectively, equations (\ref{eq:3.6}) and (\ref{eq:3.8}), we obtain:

\begin{equation} \begin{split} \label{eq:5.9}
    \omega'(r,\theta)=&\frac{1}{2A^2r^3}[\hspace{0.1cm} 4A^2Ber^2+4Ajr(e^2-2mr)+B^3e((2m-r)r+e^2(A^2r^2-1))\hspace{0.1cm}]+\\&+\frac{(A^2r^2-1)(e^2+r(r-2m))(4Ajr-B^3(e+2Aer\cos\theta))}{2A^2r^3(1+Ar\cos\theta)^2}
\end{split} \end{equation}

\begin{equation} \label{eq:5.10}
    A_t'(r,\theta)=S(r,\theta)-\omega(r,\theta)A_\phi(r,\theta)
\end{equation}

with the auxiliary function $S(r,\theta)$:

\begin{equation} \begin{split} \label{eq:5.11}
    S(r,\theta)=&\frac{e}{r}+\frac{e}{4r(1+Ar\cos\theta)^2}3B^2[\hspace{0.1cm}-2Ar^3\cos\theta-e^2\cos^2\theta+2mr(1+2Ar\cos\theta+\\&+\cos^2\theta)+r^2((A^2e^2-1)\cos^2\theta-1)\hspace{0.1cm}]
\end{split} \end{equation}

Despite the rather long and complicated expression for $\omega'(r,\theta)$ we can clearly see the Ehlers and Harrison individual contributions to the rotation, just like in the previous case (\ref{eq:4.7}) (which is precisely (\ref{eq:5.9})'s $A\xrightarrow{}0$ limit). We also notice a coupling between the acceleration parameter A and both transformation parameters j and B.

We now have all the functions we need in order to write the new metric:

\begin{equation} \label{eq:5.12}
        ds^2=-\frac{f(r,\theta)}{|\Lambda(r,\theta)|^2}(d\phi-\omega'(r,\theta)dt)^2+\frac{|\Lambda(r,\theta)|^2}{f(r,\theta)}\left(\rho(r,\theta)^2dt^2-\frac{r^4\sin^2\theta P(\theta)}{(1+Ar\cos\theta)^2}\left(\frac{dr^2}{Q(r)}-\frac{d\theta^2}{P(\theta)} \right) \right)
\end{equation}

\subsection{Physical and geometrical properties}
As we can see, (\ref{eq:5.12}) is a five parameter metric: the mass $m$, the electric charge of the seed $e$, the magnetic field of the background $B$, the angular velocity of the background $j$ and the acceleration $A$ of the black hole. 
This spacetime may be viewed as a non-asymptotically flat, non-diagonal deformation of the seed metric, i.e. as a charged C-metric\footnote{A charged C-metric represents an accelerating Reissner-Nordström black hole.} inserted in a swirling, magnetic universe. Let us note that, by construction, (\ref{eq:5.12}) is a solution to the Einstein-Maxwell field equations, but this has also been verified on the Wolfram Mathematica software.

We will now examine some physical and geometrical properties of the accelerated Reissner-Nordström black hole embedded in a swirling, magnetic universe.

\subsubsection[Analysing the spacetime's frame-dragging]{\large{Analysing the spacetime's frame-dragging}}

Let us observe the behaviour of the angular velocity $\Omega(r,\theta)$ along the z axis, in particular where $\theta=0$ and $\theta=\pi$:

\begin{equation*} \begin{split}
     \Omega|_{\theta=0}=\lim_{\theta\to 0}\left(-\frac{g_{t\phi}}{g_{\phi\phi}}\right)=&-\frac{A^2Be(4+B^2e^2)r-4jr+A(4Be+4j(2e^2-4mr+r^2))}{2Ar(1+Ar)}+\\&+\frac{B^3e(e^2+2r(r-2m))}{2r(1+Ar)}
\end{split} \end{equation*}

\begin{equation*} \begin{split}
     \Omega|_{\theta=0}=\lim_{\theta\to 0}\left(-\frac{g_{t\phi}}{g_{\phi\phi}}\right)=&-\frac{A^2Be(4+B^2e^2)r+4jr+A(-4Be+4j(2e^2-4mr+r^2))}{2Ar(Ar-1)}+\\&-\frac{B^3e(e^2+2r(r-2m))}{2r(Ar-1)}
\end{split} \end{equation*}

Just like in the non-accelerating case, $\Omega(r,\theta)$ is not constant along the z axis, but rather has two different behaviours for $\theta=0$ and $\theta=\pi$. As we mentioned, this is typical of spacetimes such as magnetised Kerr solutions but does not occur in asymptotically flat spacetimes.

As in the non-accelerating case, the metric presents a frame-dragging phenomenon (explained in Section 4.2.1). This is given by:

\begin{equation*}
    \frac{d\phi}{dt}=-\frac{g_{t\phi}}{g_{\phi\phi}}=-\omega'(r,\theta)
\end{equation*}

where $\omega'(r,\theta)$ is our metric function (\ref{eq:5.9}).

Considering now the $r\xrightarrow{}\infty$ limit of the angular velocity, we have:

\begin{equation*}
    \lim_{r\to \infty}\frac{d\phi}{dt}=\frac{B^3e\cos\theta-2j}{A\cos^2\theta}
\end{equation*}

Unlike the non-accelerating case, it does not grow unbounded in the $r\xrightarrow{}\infty$ limit. Therefore the superluminal observers paradox doesn't even present itself, so there is no violation of causality and no closed timelike curves.

\subsubsection[Dirac and Misner strings]{\large{Misner and Dirac strings}}

A brief explanation of Dirac and Misner strings can be found in Section 4.2.2, where we first mentioned them. We can once again check if they exist in our spacetime and, in case they do, if they are removable by some appropriate parameter choice.

The presence of Dirac strings is a mathematical consequence of the inequality of these two limits:

\[\lim_{\theta\to 0} A_\phi = \frac{2e^2(12B+B^3e^2+8ej)}{16+24B^2e^2+B^4e^4+32Be^3j+16e^4j^2} \]

\[\lim_{\theta\to \pi} A_\phi = \frac{2e^2(12B+B^3e^2-8ej)}{16+24B^2e^2+B^4e^4-32Be^3j+16e^4j^2} \]

These two limits not equal and we notice that the acceleration parameter A does not appear in this particular limit of $A_\phi$. What is more surprising is that the values we found are precisely (\ref{eq:dir1}) and (\ref{eq:dir2}), $A_\phi$'s $\theta\xrightarrow{}0$ and $\theta\xrightarrow{}\pi$ limits in the non accelerating case. Therefore, we conclude that the Dirac string can be removed by taking the following value of the rotational parameter $j$:

\begin{equation*}
    j=\pm j_D=\pm\frac{\sqrt{24B^2e^2+3B^4e^4-16}}{4e^2}
\end{equation*}

Moving on to Misner strings: we can check for these by analysing $\omega_{el}(r,\theta)$ as it appears in the electric LWP ansatz. We do so by evaluating the following limits:

\[\lim_{\theta\to 0} \frac{g_{t\phi}}{g_{tt}} \hspace{0.2cm},\hspace{0.3cm} \lim_{\theta\to\pi} \frac{g_{t\phi}}{g_{tt}} \]

where $g_{t\phi}/g_{tt}$ represents $\omega_{el}(r,\theta)$. As we also found in the non-accelerating case, both limits are equal to zero, which means that our metric does not present any Misner strings.

\subsubsection[Conical singularities]{\large{Conical singularities}}

The main reason why we introduced the accelerating Reissner-Nordström spacetime as a new seed was to attempt to remove the conical singularities that plagued the previous metric. We will now follow the same procedure of Section 4.2.3 in order to look for these angular defects or lack thereof.

We consider the ratio between the circumference and the radius of a small circle around the z-axis, for $\theta=0$ and $\theta=\pi$. If no angular defects are present, the ratio must be equal to $2\pi$. This is equivalent to the following condition:

\[ \lim_{\theta\to 0} \frac{1}{\theta} \int_{0}^{2\pi}\sqrt{\frac{g_{\phi\phi}}{g_{\theta\theta}}} \,d\phi = 2\pi = \lim_{\theta\to \pi} \frac{1}{\pi-\theta} \int_{0}^{2\pi}\sqrt{\frac{g_{\phi\phi}}{g_{\theta\theta}}} \,d\phi \]

As previously done, $\phi$ is redefined as $\phi'=\phi\cdot\Delta\phi$ so that its range is now [$0,2\pi\Delta\phi$].

The resulting limits are:

\begin{equation} \label{eq:5.13}
    \lim_{\theta\to 0} \frac{1}{\theta} \int_{0}^{2\pi}\sqrt{\frac{g_{\phi\phi}}{g_{\theta\theta}}} \,d\phi = \frac{16(1+A^2e^2+2Am)}{16+24B^2e^2+B^4e^4+32Be^3j+16e^4j^2}2\pi\Delta\phi
\end{equation}

\begin{equation} \label{eq:5.14}
    \lim_{\theta\to\pi} \frac{1}{\pi-\theta} \int_{0}^{2\pi}\sqrt{\frac{g_{\phi\phi}}{g_{\theta\theta}}} \,d\phi = \frac{16(1+A^2e^2-2Am)}{16+24B^2e^2+B^4e^4-32Be^3j+16e^4j^2}2\pi\Delta\phi
\end{equation}

In order to remove the conical singularity in $\theta=0$, $\Delta\phi$ is set to:

\begin{equation*}                 \Delta\phi=\frac{16+24B^2e^2+B^4e^4+32Be^3j+16e^4j^2}{16(1+A^2e^2+2Am)}
\end{equation*}

And to regularise the conical singularity in $\theta=\pi$, the following condition must be met:

\begin{equation} \label{eq:5.15}              \frac{16(1+A^2e^2-2Am)}{16+24B^2e^2+B^4e^4-32Be^3j+16e^4j^2}\cdot\frac{16+24B^2e^2+B^4e^4+32Be^3j+16e^4j^2}{16(1+A^2e^2+2Am)}=1
\end{equation}

Clearly, if $A$ were zero and the spin-spin coupling $Be$-$j$ were not present, this condition would always be met. Furthermore, if this condition is satisfied, the force necessary to accelerate the black hole is provided by the spin-spin interaction and it is no longer necessary to introduce the hypothesis of cosmic strings or struts.

At low energies, i.e. at low values of $A$, $B$ and $j$, the condition translates to

\begin{equation*}                 
mA\approx B e^3 j
\end{equation*}

clearly showing that the nature of the force is a spin-spin interaction between the rotation caused by the Lorentz force and the one caused by the swirling background. 

In particular, (\ref{eq:5.15}) sets the acceleration parameter A to

\begin{equation*}                 
A=\frac{(16 + 24 B^2 e^2 + B^4 e^4 + 
    16 e^4 j^2)m \pm \sqrt{(16 + 24 B^2 e^2 + 
     B^4 e^4 + 16 e^4 j^2)^2 m^2-1024 B^2 e^8 j^2}}{32 B e^5 j}
\end{equation*}

However, for a fully regularised spacetime, magnetic monopole related singularities must be removed as well. This leads to inserting $\pm j_D$, i.e. the values of $j$ that remove the Dirac string, in the previous expression. For instance, inserting $+j_D$ yields:

\begin{equation*}                 
A=\frac{12 B m + B^3 e^2 m \pm \sqrt{64 + B^2 (-12 e^2 (8 + B^2 e^2) + (12 + B^2 e^2)^2 m^2)}
 }{2 e \sqrt{24 B^2 e^2 + 3 B^4 e^4-16}}
\end{equation*}

It can be easily seen that at low energies, i.e. low values of $B$, these values of $A$ become imaginary. This means that the condition for the conical singularity removal is never met and therefore, at low energies, the spacetime can never be fully regular because either a Dirac string or a conical singularity will be present.

\clearpage

\addcontentsline{toc}{section}{Conclusions}
\section*{\Huge{Conclusions}\markboth{Conclusions}{Conclusions}}
This work was devoted to show the effect of the Harrison-Ehlers transformation on a well-known solution of the Einstein-Maxwell equations, and, more in general, the power and simplicity of the Ernst generating technique. A new solution of the Einstein-Maxwell equations and its accelerated generalisation were found and analysed.

The non accelerated solution was found to be a Reissner-Nordström black hole immersed in a swirling, magnetic universe. An ergoregion analysis was carried out for this spacetime and its two subcases, the swirling and the magnetic Reissner-Nordström spacetimes. Unlike what was found in the two subcases, the complete spacetime's ergoregions are asymmetrical due to the interaction of the two individual contributions to its rotation. Due to the nature of the two contributions, i.e. one causing uniform rotation and the other causing a whirlpool rotation in two opposite directions, the complete spacetime must rotate in a highly non-uniform manner, meaning that the frame dragging is higher in some areas compared to others. Moreover, we saw that this spacetime was plagued not only with conical singularities but also with Dirac strings, while Misner strings and closed timelike curves were not present. In particular, Dirac strings were found to be removable by an appropriate choice of the background rotation parameter, while that was not the case for conical singularities. For this reason, in an attempt to fully regularise the metric outside the event horizons, the accelerated Reissner-Nordström black hole was introduced as a new ``seed''.

The spacetime resulting from the accelerated Reissner-Nordström black hole, i.e. charged C-metric, showed some interesting properties. Much like the previous case, Misner strings and causality violations were absent, unlike Dirac strings and conical singularities. It is interesting to note that the condition for the removal of Dirac strings is precisely the one found in the non accelerated case. The condition for the removal of conical singularities regarded the acceleration parameter and it was applied in conjunction with the Dirac string condition. This union was found to work just fine at regular and high energies, but at low energies (at low values of the background magnetic field) the two conditions were found to be incompatible. This leads to the realisation that, at low energies, this spacetime can never be fully regular outside the event horizons, because it will either be affected by a conical singularity or by a Dirac string. 
\clearpage

\newpage
\mbox{}
\thispagestyle{plain}
\newpage

\addcontentsline{toc}{section}{Appendix A}
\appendix
\section*{{\centering\normalfont\Large\bfseries
  {\textit{Appendix A}}}
\hrule
\vspace{1ex}
\Huge{C-metrics}}\markboth{Appendix A - C-metrics}{Appendix A - C-metrics}\label{appendix_a}
In this appendix, we will focus on C-metrics and their general properties to the extent that is relevant to this thesis, however a more extensive approach can be found in \cite{Cmetric}.

The name C-metric was first coined in \cite{Witten}, while Kinnersley and Walker (\cite{kinnwalk}) showed that this metric can be regarded as a black hole that accelerates because of the presence of cosmic strings (or struts), which mathematically correspond to a conical singularity.

A first expression of the C-metric line element is:

\begin{equation*}
    ds^2=\frac{1}{A^2(x+y)^2}\left(-Fdt^2+\frac{dy^2}{F}+\frac{dx^2}{G}+Gd\phi^2 \right)
\end{equation*}

where F and G are cubic functions of their respective variables:

\begin{equation*}
    F(y)=y^2-1-2MAy^3
\end{equation*}

\begin{equation*}
    G(x)=1-x^2-2MAx^3
\end{equation*}

We have two constant parameters $M$ and $A$ and interestingly this metric has no obvious limit for $A\xrightarrow{}0$ but reduces to the Minkowski spacetime when $M=0$.

A significant simplification to this metric's structure was achieved by Hong and Teo in the 2003 article \cite{Hong}, but such details are not relevant to us. Instead, let us immediately rewrite the C-metric in a form of spherical coordinates through the transformation\footnote{This coordinate transformation comes quite naturally since x is assumed to be in the range [-1,1], details are found in \cite{Cmetric}.}:

\begin{equation*} \begin{cases}
    x=cos\theta \\
    y=\frac{1}{\alpha r}
\end{cases} \end{equation*}

The expression for the metric is then:

\begin{equation*}
    ds^2=\frac{1}{(1+\alpha rcos\theta)^2}\left(-Q(r)dt^2+\frac{dr^2}{Q(r)}+\frac{r^2d\theta^2}{P(\theta)}+P(\theta)r^2sin^2\theta d\phi^2 \right)
\end{equation*}

where $\theta\in[0,\pi]$, $\phi\in[0,2\pi C]$\footnote{The constant C and its presence in $\phi$'s range will be explained later.} and

\begin{equation*} \begin{cases}
    P(\theta)=1+2\alpha m\cos\theta \\
    Q(r)=(1-\alpha^2r^2)(1-2m/r)
\end{cases} \end{equation*}

The rescaling of the parameters $M\xrightarrow{}m$ and $A\xrightarrow{}\alpha$ makes the following possible: the C-metric, in these coordinates, reduces exactly to the Schwarzschild spacetime when $\alpha=0$. Since this metric is indeed a one-parameter generalisation of the Schwarzschild solution, we can keep thinking of $m$ as the source's mass parameter. In addition, we will interpret $\alpha$ as the acceleration parameter.

One peculiarity of this solution is that, while it maintains the traditional (Schwarzschild) event horizon at $r=2m$, it also presents another singularity at $r=1/\alpha$, due to Q's expression.

In order to investigate this spacetime's physical properties, we examine its symmetry axis to check for conical singularities, precisely as we did in Section 4.2.3. Around the half axis $\theta=0$ we find:

\begin{equation*}
    \lim_{\theta\to 0} \frac{1}{\theta} \int_{0}^{2\pi}\sqrt{\frac{g_{\phi\phi}}{g_{\theta\theta}}} \,d\phi = \lim_{\theta\to 0}2\pi C P(\theta)=\lim_{\theta\to 0}2\pi C(1+2\alpha m)
\end{equation*}

and for  $\theta=\pi$ we find:

\begin{equation*}
    \lim_{\theta\to \pi} \frac{1}{\pi-\theta} \int_{0}^{2\pi}\sqrt{\frac{g_{\phi\phi}}{g_{\theta\theta}}} \,d\phi = \lim_{\theta\to 0}2\pi C P(\theta)=\lim_{\theta\to 0}2\pi C(1-2\alpha m)
\end{equation*}

Unless $\alpha m=0$, we have a conical singularity with different conicity. As we saw in Section 4.2.3, we can remove one deficit or excess angle\footnote{We interpret deficit angles as pulling strings and excess angles as pushing struts.} with a good choice of C, but not both at once. If we choose to remove the conical singularity at $\theta=0$ by setting

\begin{equation*}
    C=\frac{1}{1+2\alpha m}
\end{equation*}

our deficit angles become:

\begin{equation*} \begin{cases}
    \delta_0=2\pi-2\pi=0 \\
    \delta_\pi=2\pi-2\pi\hspace{0.1cm}\cfrac{1-2\alpha m}{1+2\alpha m}=\cfrac{8\pi\alpha m}{1+2\alpha m}
\end{cases} \end{equation*}

As we already mentioned, the remaining conical singularity, with constant deficit angle, on the half axis $\theta=\pi$ can be viewed as a semi-infinite cosmic string under tension.

\clearpage

\newpage
\mbox{}
\thispagestyle{plain}
\newpage

\addcontentsline{toc}{section}{Bibliography}
\bibliography{references}

\newpage
\mbox{}
\thispagestyle{plain}
\newpage

\addcontentsline{toc}{section}{Acknowledgements}
\section*{\Huge{Acknowledgements}\markboth{Acknowledgements}{Acknowledgements}}
Un primo sentito ringraziamento va naturalmente al Dottor Astorino e alla Professoressa Klemm per avermi proposto questo tema di ricerca di tesi, per aver seguito il mio lavoro e per avermi dato preziosi consigli e ispirazione per il mio cammino.

In secondo luogo, ringrazio la mia famiglia, in particolare i miei genitori Roberta e Francesco, i miei nonni Sira, Guido e Anna e mia sorella Ludovica. Hanno sempre creduto in me, dandomi fiducia e supporto per tutti questi anni.

Per concludere, vorrei ringraziare Francesco e Aurelio, indispensabili compagni di questo viaggio di tre anni, e Pietro, che mi ha sempre incoraggiata con un po' di sana competizione.
A tutti gli altri amici un sincero grazie per aver donato un po' di leggerezza a questi anni impegnativi.

\clearpage

\end{document}